\documentclass[twocolumn,pre]{revtex4}   

\usepackage{dcolumn}
\usepackage{amsmath}
\usepackage{calrsfs}

\usepackage{graphicx}
\usepackage{subfigure}

\newcommand{\e}{\mathrm{e}}
\newcommand{\half}{\tfrac12}
\newcommand{\set}[1]{\lbrace#1\rbrace}
\newcommand{\etal}{{\it{}et~al.}}
\newcommand{\defn}{\emph}

\newcommand{\argmax}{\mathop\mathrm{argmax}}

\newcommand\cin{c_\textrm{in}}
\newcommand\cout{c_\textrm{out}}

\begin{document}

\title{Community detection in networks with unequal groups}
\author{Pan Zhang,$^1$ Cristopher Moore,$^1$ and M. E. J. Newman$^{2,1}$}

\affiliation{
$^{1}$Santa Fe Institute, Santa Fe, New Mexico 87501, USA \\
$^{2}$Physics Department and the Center for the Study of Complex Systems, University of Michigan, Ann Arbor, MI 48109, USA}

\begin{abstract}
  Recently, a phase transition has been discovered in the network community detection problem below which no algorithm can tell which nodes belong to which communities with success any better than a random guess.  This result has, however, so far been limited to the case where the communities have the same size or the same average degree.  Here we consider the case where the sizes or average degrees are different.  This asymmetry allows us to assign nodes to communities with better-than-random success by examining their local neighborhoods.  Using the cavity method, we show that this removes the detectability transition completely for networks with four groups or fewer, while for more than four groups the transition persists up to a critical amount of asymmetry but not beyond.  The critical point in the latter case coincides with the point at which local information percolates, causing a global transition from a less-accurate solution to a more-accurate one.
\end{abstract}
\maketitle

\section{Introduction}

Community detection, the division of a network into well-connected groups of nodes with only sparser connections between groups, has been the subject of vigorous research in a number of fields including physics, statistics, and computer science~\cite{Fortunato10}.  A string of recent discoveries, however, have revealed that there are fundamental limits to our ability to detect community structure~\cite{RL08,DKMZ11a,DKMZ11b,HRN12,mossel2012reconstruction,mossel-colt,Massoulie2013}.  Using techniques from statistical physics and probability theory, it has been shown that there can exist networks that possess underlying community structure and yet that structure is undetectable.  In particular, for certain classes of model networks it has been shown that there exists a sharp \defn{detectability threshold} above which efficient algorithms for community detection exist, but below which no algorithm of any kind can classify nodes into their correct communities with success any better than a random guess---or even detect the existence of communities in the network---if given only the network topology as input.

The simplest demonstration of this effect makes use of the \defn{stochastic block model}, a probabilistic generative network model that allows one to create artificial networks with any number of communities of any size~\cite{HLL83}.  For networks generated using this model the existence and location of the detectability transition has been rigorously proven for the case of two communities of equal size~\cite{mossel2012reconstruction,mossel-colt,Massoulie2013}.  The transition is a continuous one, with the fraction of correctly classified nodes playing the role of order parameter.  When the number of groups is increased, the phase transition becomes more complicated, analogous to that of random constraint satisfaction problems~\cite{Krzakala2007a}.  For five or more groups (or four or more in the disassortative or antiferromagnetic case) there is a ``hard/easy'' threshold where the accuracy achievable by an efficient algorithm undergoes a first-order transition and jumps discontinuously.  Immediately below this point there is a regime where community detection is possible in principle, but is believed to require exponential time~\cite{DKMZ11a,DKMZ11b}.

These results are for the symmetric case where the groups have equal size or, more generally, equal average degree.  In this case, every node has the same probability distribution of local neighborhoods, so that the local environment of a node gives us no information about what community it belongs to.  In this paper we investigate the less well-studied case where the groups have different sizes or average degrees, which is of obvious relevance to real networks.  This case is harder to analyze than the case of equal groups.  We tackle it using two approaches, both based on the cavity method of~\cite{DKMZ11a,DKMZ11b}.  In the first, we perform a perturbative expansion of the cavity method equations; in the second we consider the behavior of the equations under finite iteration.

It is straightforward to see that having unequal groups makes community detection easier.  When different groups have different average degrees we can use the node degree as a simple proxy for group membership.  And making the group sizes unequal in general makes the average degrees unequal too (as we will show), so again we can use degree as a proxy.  Furthermore, by propagating degree-based estimates of group membership through the network using a message-passing (belief propagation) algorithm, we can improve on the accuracy of this initial classification, labeling nodes based not only on their own degrees but also on the degrees of their neighbors, their neighbors' neighbors, and so on.  Iterating the message-passing calculation repeatedly corresponds to increasing the radius of the network neighborhood from which we draw information, until the classification reaches a fixed point when all information has been taken into account.

It is known that the classification provided by this fixed point (or if there are multiple fixed points, the one with the highest likelihood or lowest Bethe free energy---see below) is optimal, in the sense that no other algorithm for community detection can do a better job~\cite{DKMZ11a}.  In particular, if the fixed point does a poor job of assigning nodes to groups---or if it fails completely---then no other method will return better performance and it is this observation that allows us to say when the structure in the network becomes undetectable.

Using these methods, we show in this paper that for four or fewer groups the second-order detectability transition of the equal-groups case disappears, but that for five or more groups the first-order transition, and the coexistence regime where several competing fixed points exist, persist up to a critical level of asymmetry.  In all cases we can classify the nodes better than chance, no matter what the parameter values are, but while in some cases our final accuracy is a smooth function of the parameters, in others there is a sudden jump from low accuracy based on purely local information to high accuracy based on propagating information globally across the network.

We note that this phenomenology is qualitatively similar to the case of ``semisupervised'' community detection, where we are given the true labels of a small fraction of nodes~\cite{Zhang2014phase,Steeg2013}, and also to the Franz--Parisi spin-glass model~\cite{Franz1997}, where each node has an external field pointing it to a reference state.  In these models, the known labels or external fields break the symmetry and provide local information which propagates under belief propagation, causing the coexistence region to shrink and finally disappear at a critical point.  However, the scenario we study here is different in that our local information comes directly from the topology of the network itself, without the need for any ``metadata'' or external field.

In Sections~\ref{sec:model} and~\ref{sec:detectability} we define the stochastic block model and describe in detail previous results on detectability and how they were reached.  Then in Section~\ref{sec:unequal} we develop the theory for networks with groups of unequal size and degree, including series expansions around the limit of weak structure and optimal local classifiers based on neighborhoods of a given radius.  In Section~\ref{sec:numerics} we present extensive numerical tests on the stochastic block model that confirm the picture painted by our theoretical results.  In Section~\ref{sec:conclusions} we give our conclusions.

\section{The stochastic block model}
\label{sec:model}

The stochastic block model is a model for networks containing community structure.  It can be used both in a forward direction for generating artificial networks with tunable structure and in reverse for detecting the presence of communities in network data by fitting the model to the data.  In this paper we do both: we use the model to generate test networks with known community structure, and then attempt to detect that structure by fitting that same model to the network.  This dual approach is central to understanding when community structure is or is not detectable, since there is no better way to detect the structure in a network (or any other data set) than to fit it to the very model used to generate that structure in the first place.  As pointed out by Decelle~\etal~\cite{DKMZ11a}, this means that if we fail to detect the community structure in our networks by this method, then all other methods must also fail on the same networks.  The structure in such networks can thus fairly be said to be undetectable.

The definition of the stochastic block model is as follows.  Each of $n$ nodes is assigned to one of $q$ groups, with probabilities~$\gamma_1,\ldots,\gamma_q$ of assignment to group 1 to~$q$ respectively.  Thus $\gamma_a$ is the expected size of group~$a$ as a fraction of~$n$.  Once the group assignments are chosen, edges are placed between node pairs independently at random with probabilities~$p_{ab}$ that depend only on the groups $a,b$ that a pair belongs to.  If the diagonal elements~$p_{aa}$ of the matrix of probabilities are larger than the off-diagonal ones, the resulting network will have traditional ``assortative'' community structure in which edges are more probable within groups than between them.  However, other types of structure are possible and are observed in certain real-world networks, including ``disassortative'' structure where edges are more common between groups than within them, or mixed structures in which different groups may be variously assortative or disassortative with respect to one another.

In this paper we focus on the case of a sparse network with constant
average node degree in the limit of large network size, meaning that the
edge probabilities~$p_{ab}$ scale as~$1/n$.  Specifically we set $p_{ab} =
c_{ab}/n$ where the $c_{ab}$ are constants.  Then the expected degree~$c_a$ of
a node in group~$a$ is the sum of the probabilities of connection between
it and all other nodes, averaged over all possible assignments of nodes to
communities.  Letting $s_i$ denote the group to which node $i$ belongs, we
have
\begin{align}
c_a &= \sum_{\set{s_i}} \prod_i \gamma_{s_i} \sum_i p_{a,s_i}
  = \sum_i \sum_b p_{ab} \gamma_b
  = n \sum_b {c_{ab}\over n} \gamma_b \nonumber\\
  &= \sum_b c_{ab} \gamma_b.
\label{eq:avgc0}
\end{align}
The sparse case appears to be representative of most real-world networks and  also displays a richer phase transition structure in the community detection problem.

\subsection{Fitting the stochastic block model\\to network data}
\label{sec:bp}

In this paper we consider the following problem.  An undirected network is generated by the stochastic block model for some choice of~$\{\gamma_a\}$ and~$\{c_{ab}\}$, and our goal is to find the best fit of the same model to the network data, so as to recover the community assignments planted in the network.

In performing the fit, we will assume that the values of the parameters $\gamma_a$ and $c_{ab}$ used to generate the network are known exactly.  The only quantities we need to determine by our fit are which nodes belong to which groups.  This is a somewhat unrealistic assumption.  In general, nothing is known beforehand, and one must learn the values of the parameters as well as the group assignments.  In some cases we can do this using an expectation--maximization algorithm~\cite{DLR77,NS01,MK08,DKMZ11a,DKMZ11b}.  However, our goal here is to understand the fundamental limits on our ability to detect community structure and for this purpose the simpler setup considered here is a useful one.  If it is impossible to detect community structure when we are given the values of the parameters, then it will still be impossible when we are not given them.  Hence the accuracy we can achieve given the parameter values sets an upper bound on what we can achieve when the parameters are unknown.

Given the parameters $\{\gamma_a\}$ and~$\{c_{ab}\}$, the optimal group assignments can be calculated by maximizing the likelihood that the observed network was generated by the model.  In the case of sparse networks it can be misleading to focus only on the single assignment that maximizes the likelihood, which can result in overfitting of the data.  Instead we focus on the posterior distribution $\mu(\set{s_i})$ over group assignments, and especially the \emph{marginal probability} of group membership for each node, i.e.,~the probability $\mu_a^i$ that node $i$ belongs to group $a$:
\begin{equation}
\mu_a^i = \sum_{\set{s_i}} \mu(\set{s_i}) \,\delta_{a,s_i},
\label{eq:onenode}
\end{equation}
where $\delta_{a,b}$ is the Kronecker delta.  In particular, if our goal is to maximize the fraction of nodes labeled correctly, the optimal strategy is to label each node $i$ with its most-likely group, given by $\argmax_a \mu_a^i$.

The optimal (maximum-likelihood) value of the posterior distribution can be shown (via a standard derivation involving Jensen's inequality) to be given by maximizing the quantity
\begin{align}
\mathcal{L} &= \sum_a \sum_i \mu_a^i \log \gamma_a
  + \sum_{ab} \sum_{(i,j)} \mu_{ab}^{ij} \log c_{ab} \nonumber\\
  &\qquad{} - {1\over n} \sum_{ab} \sum_{ij} \mu_{ab}^{ij} c_{ab}
      - \sum_{\set{s_i}} \mu(\set{s_i}) \log \mu(\set{s_i}) 
\label{eq:ll}
\end{align}
as a function of the distribution $\mu(\set{s_i})$.  
Here the notation $\sum_{(i,j)}$ denotes a sum over all edges~$(i,j)$ in the network, and $\mu_{ab}^{ij}$ is the two-node marginal probability that nodes~$i$ and~$j$ belong to groups~$a$ and~$b$ respectively:
\begin{equation}
\mu_{ab}^{ij} = \sum_{\set{s_i}} \mu(\set{s_i}) \,\delta_{a,s_i} \,\delta_{b,s_j}.
\label{eq:twonode}
\end{equation}

The quantity $\mathcal{L}$ has the character of a free energy.  Its maximization requires us to find a distribution $\mu$ whose one- and two-node marginals give a large value for the average log-likelihood of the observed network (the first three terms in~$\mathcal{L}$), while also giving a large value for the entropy term $-\sum_{\set{s_i}} \mu(\set{s_i}) \log \mu(\set{s_i})$.  The traditional approach to this problem, borrowed directly from statistical mechanics, is to treat $\mu(\set{s_i})$ as a Gibbs distribution over ``states''~$\set{s_i}$ whose Hamiltonian consists of (minus) the first three terms in~$\mathcal{L}$ (the ``internal energy'') and sample from this distribution using a Monte Carlo algorithm.  
However, obtaining good statistics on the marginals requires us to take many independent samples, which is computationally expensive.

An elegant alternative, better suited to our current aims, is the belief propagation method proposed recently by Decelle~\etal~\cite{DKMZ11a}.  Belief propagation focuses on the ``belief'' or ``message''~$\mu_a^{i\to j}$, which is an estimate of the probability that node~$i$ would belong to group~$a$ if node~$j$ were removed from the network (or, more precisely, if we lacked information about whether or not $i$ and $j$ have an edge between them).  The removal of a node corresponds to the cavity method of statistical mechanics: it allows us to write down a set of self-consistent equations that must be satisfied by the beliefs thus~\cite{DKMZ11a}:
\begin{equation}
\mu_a^{i\to j} = {\gamma_a\over Z_{i\to j}}
   \exp \biggl( - {1\over n} \sum_k \sum_b c_{ab} \mu_b^k \biggr)
   \prod_{k \in \partial i\backslash j}
   \sum_b c_{ab} \mu_b^{k\to i}.
\label{eq:bp}
\end{equation}
Here $\partial i$ denotes the set of neighbors of node~$i$ and 
$\partial i\backslash j$ denotes that set exclusive of node~$j$.  
The quantity~$Z_{i\to j}$ is a normalizing constant that ensures that 
$\sum_a \mu_a^{i \to j} = 1$:
\begin{equation}
Z_{i\to j} = \sum_a \gamma_a
   \exp \biggl( - {1\over n} \sum_k \sum_b c_{ab} \mu_b^k \biggr)
   \prod_{k \in \partial i\backslash j}
   \sum_b c_{ab} \mu_b^{k\to i}.
\label{eq:Zitoj}
\end{equation}

These equations assume that $i$'s neighbors are independent of each other given its state $s_i$, or equivalently, that $i$'s neighbors are correlated only through their interaction with $i$.  As a result, belief propagation is only exact on trees; on a finite graph with loops, it is merely an approximation.  As long as correlations in the network decay with distance, however, it becomes exact in the limit of large size for a network that is ``locally treelike,'' meaning that almost all vertices have neighborhoods which are trees up to a radius of $O(\log n)$.  Networks generated by the stochastic block model satisfy this condition in the sparse case considered here, and hence we expect belief propagation to give exact results in the large-$n$ limit.

Implementing belief propagation consists of solving Eq.~\eqref{eq:bp} by simple iteration starting from an appropriate initial condition and iterating until the beliefs converge to a fixed point.  The one-node marginal probabilities~$\mu_a^i$ can 
be calculated directly from the beliefs according to
\begin{equation}
\mu_a^i = {\gamma_a\over Z_i}
   \exp \biggl( - {1\over n} \sum_k \sum_b c_{ab} \mu_b^k \biggr)
   \prod_{k \in \partial i} \sum_b c_{ab} \mu_b^{k\to i},
\label{eq:bpmu}
\end{equation}
where $Z_i$ is a normalizing constant,
\begin{equation}
Z_i = \sum_ a \gamma_a
   \exp \biggl( - {1\over n} \sum_k \sum_b c_{ab} \mu_b^k \biggr)
   \prod_{k \in \partial i} \sum_b c_{ab} \mu_b^{k\to i}.
\label{eq:Zi}
\end{equation}
The two-node marginals of Eq.~\eqref{eq:twonode} can also be calculated from the beliefs.  For pairs $i, j$ connected by an edge, 
\begin{equation}
\mu_{ab}^{ij} = \frac{1}{Z_{ij}} \,c_{ab} \mu_a^{i\to j} \mu_b^{j\to i} 
\label{eq:bpmu2}
\end{equation}
where $Z_{ij}$ is another normalizing
constant:
\begin{equation}
Z_{ij} = \sum_{ab} c_{ab} \mu_a^{i\to j} \mu_b^{j\to i}.
\label{eq:Zij}
\end{equation}
In the sparse case, we can assume that pairs $i, j$ not connected by an edge are independent, so that 
\begin{equation}
\mu_{ab}^{ij} = \mu_a^i \mu_b^j .
\label{eq:bpmu1}
\end{equation}

To calculate the value of~$\mathcal{L}$ itself, we can substitute the converged values of the one- and two-node marginals obtained from the belief propagation equations~\eqref{eq:bpmu2} and~\eqref{eq:bpmu1} back into the log-likelihood, Eq.~\eqref{eq:ll}.  The final entropy term in~\eqref{eq:ll} requires an expression for the full joint posterior distribution~$\mu(\set{s_i})$, which we assume takes the factorized form
\begin{equation}
\mu(\set{s_i}) = {\prod_{(i,j)} \mu_{s_is_j}^{ij}\over
          \prod_i \bigl( \mu_{s_i}^i \bigr)^{d_i-1}} , 
\label{eq:bethe}
\end{equation}
where $d_i$ is the degree of node~$i$.  (Again, this form is exact on trees, and asymptotically exact on locally treelike networks in the limit of large size; on finite networks with loops it is only approximate, and indeed does not even sum to~1.)  After some manipulation, one can then show that the converged value of~$\mathcal{L}$, which is also equal to the log-likelihood, is
\begin{equation}
\mathcal{L} = \sum_{(i,j)} \log Z_{ij} - \sum_i \log Z_i
              + {1\over n} \sum_{ab} c_{ab} \sum_i \mu_a^i \sum_j \mu_b^j,
\end{equation}
with $Z_i$ and $Z_{ij}$ as in Eqs.~\eqref{eq:Zi} and~\eqref{eq:Zij}.  This quantity (or, rather, minus this quantity) is called the \emph{Bethe free energy}, and it can be shown~\cite{Yedidia2001,Yedidia2005} that fixed points of belief propagation are stationary points of the Bethe free energy.  In particular, there is a stable fixed point that maximizes $\mathcal{L}$ whenever $\mu$ takes the form~\eqref{eq:bethe}.  However, belief propagation often has many fixed points in addition to this one, so it is possible for it to converge to a local optimum of $\mathcal{L}$ rather than the required global optimum.  To get around this problem one typically runs the belief propagation calculation multiple times with different initial conditions and selects, from the fixed points found, the one with the highest log-likelihood (or the lowest Bethe free energy).

In many regimes this approach works well.  However, it can also happen that the global optimum has an exponentially small basin of attraction---that is, the set of initial messages that would cause belief propagation to converge to it has exponentially small volume.  In that case, finding it can be computationally difficult, which can lead to interesting behaviors, as we will see.

\section{Detectability transitions}
\label{sec:detectability}

Belief propagation is a fast and practical method for community detection in networks and has been employed extensively to fit the stochastic block model and other related models to network data~\cite{DKMZ11a,DKMZ11b,Yan14,ZMN15,NP15}.  It is also a powerful tool for the formal analysis of algorithm performance.  By analyzing the fixed points of the belief propagation equations, Eq.~\eqref{eq:bp}, we can make statements about whether the method is, or is not, able to find the communities in a network.  And since the maximum-likelihood fit performed by the algorithm is optimal in the sense described in Section~\ref{sec:model}, if the belief propagation algorithm fails, i.e.,~if the fixed point with the highest likelihood does not give the correct communities, this implies (for locally treelike networks) that all other algorithms must also fail.  Thus results for belief propagation tell us not just about one particular algorithm, but about all possible algorithms for community detection.

Arguments of this type allowed Decelle~\etal~\cite{DKMZ11a,DKMZ11b} to show that there exist regions in the parameter space of the stochastic block model where community structure is undetectable by any means.  Specifically, they showed that if the average degrees, Eq.~\eqref{eq:avgc0}, are the same for all groups, there is a trivial fixed point where $\mu_a^{i \to j} = \mu_a^i = \gamma_a$.  If belief propagation settles at this fixed point, then it returns results no better than guessing node labels based on the prior probabilities~$\gamma_a$.  For example, in the special case where the $q$ groups have equal size $\gamma_a = 1/q$, belief propagation concludes that all nodes are equally likely to belong to all groups, and assigns nodes to groups with accuracy no better than flipping a $q$-sided coin.

The so-called \defn{hard/easy transition} corresponds to a bifurcation at which this trivial fixed point becomes unstable.  This transition is known in the spin glass literature as the de Almeida--Thouless line~\cite{Almeida1978}, and in information theory as the Kesten--Stigum transition~\cite{Kesten1966,Kesten1966a} or the robust reconstruction threshold~\cite{Mezard2006,janson2004robust}.  Above this transition, if we initialize belief propagation with random messages, or even with just a small perturbation away from the trivial fixed point, it quickly moves away from that fixed point towards another, nontrivial fixed point which is well-correlated with the true community assignment.  Thus detecting the community structure, and labeling the nodes with accuracy better than chance, is computationally easy in this regime---belief propagation succeeds at the task quickly and reliably.

The position of the hard/easy transition is relatively easy to compute in the well-studied special case where the groups have equal size and the parameters~$c_{ab}$ of the stochastic block model take just two different values:
\begin{equation}
\label{eq:cin-cout}
c_{ab} = \biggl\lbrace\begin{array}{ll}
         \cin  & \qquad\text{if $a=b$,} \\
         \cout & \qquad\text{if $a\ne b$.}
         \end{array}
\end{equation}
If $\cin$ is significantly greater than~$\cout$, this choice gives us strong assortative structure, but as $\cin$ approaches $\cout$ the structure gets weaker.  One might imagine that the structure would remain detectable, albeit with some statistical error, so long as $\cin>\cout$, but this is not the case.  Instead the trivial fixed point becomes stable when
\begin{equation}
\cin - \cout = q \sqrt{c},
\label{eq:threshold}
\end{equation}
where
\begin{equation}
c = {\cin + (q-1)\cout\over q}
\label{eq:c}
\end{equation}
is the average degree of the network as a whole.  

When the trivial fixed point is stable, belief propagation can show different behaviors depending on whether the stability is local or global, which in turn depends on the number $q$ of groups.  For $q\le4$ it is globally stable below the hard/easy transition, so that the community structure is completely undetectable (this is known rigorously for $q=2$~\cite{mossel2012reconstruction}).  Belief propagation will always converge to the trivial point and return no information about the community structure.  In this case the transition is a pitchfork bifurcation where the trivial fixed point emerges continuously from the nontrivial one.  If we define an order parameter $\sum_i \mu^i_{s_i} - 1/q$, equal to the average probability given to the correct label minus the fraction $1/q$ we would get right by chance, then this order parameter undergoes a classic second-order phase transition from a nonzero value above the critical point to zero below it.

By contrast, for $q > 4$ (or $q \ge 4$ in the disassortative case) there is a region immediately below the easy/hard transition where the trivial fixed point is locally stable, but not globally stable.  In this regime there is at least one other nontrivial fixed point that is also locally stable and corresponds to accurate classification of the nodes into their groups.  In this ``coexistence region,'' belief propagation can converge to either fixed point---and hence it may fail or succeed---depending on the initial conditions for the iteration.  Unfortunately, it appears that the basin of attraction of the accurate fixed point is exponentially small, so that we will almost always converge to the trivial fixed point if we start with random messages.  But if we have the luxury of exploring the entire space of messages, or performing an exponential number of independent runs of belief propagation, we can still find the accurate fixed point.  And if the likelihood is higher at this point than at the trivial fixed point, then the algorithm that picks the solution with higher likelihood (as described above) would choose the accurate fixed point over the trivial one and label the nodes with good accuracy.  We would, however, need to perform exponentially many runs of belief propagation to achieve this result.  Decelle~\etal~\cite{DKMZ11a,DKMZ11b} have conjectured, though it has not been proved, that in fact there exists no algorithm of any kind that will find the accurate fixed point quickly under these circumstances---specifically none that will find it in polynomial time.  If this conjecture is correct then it implies the existence of a ``hard but detectable'' regime where community detection is possible in principle but computationally hard.  (It is this regime that gives the easy/hard transition its name.)

If one continues to decrease $\cin-\cout$, there comes a point at which the likelihoods for the two fixed points cross over and the trivial fixed point becomes favored even with repeated restarts.  At this ``condensation threshold'' the system undergoes a first-order phase transition, where the fixed point that dominates the Gibbs distribution changes from the accurate one to the trivial one.  Below this point there is a ``clustered'' regime where many locally stable fixed points still exist, including the accurate one, but the algorithm that selects the solution with the highest likelihood will classify the nodes into their groups with success no better than chance.  

Thus, below the condensation point belief propagation no longer succeeds under any circumstances and the community structure becomes information-theoretically undetectable: no algorithm, even one that takes exponential time, can perform better than chance.  Finally, as we decrease $\cin-\cout$ even further, there is a ``spinodal'' or dynamical transition where the accurate fixed point disappears altogether, and the trivial point becomes globally stable.

The coexistence of more than one stable fixed point in the same parameter regime is a classic sign of a first-order phase transition.  Indeed there is a close analogy between the behavior of the community detection problem and a thermodynamic first-order transition.  As described above, one can regard the log-likelihood as (minus) the free energy of a thermodynamic system, a $q$-state Potts-like spin system in this case whose spins are the community assignments~$s_i$ of the nodes.  Fixed points of the belief propagation algorithm give us not just individual community assignments of nodes but the entire distribution~$\mu(\set{s_i})$.  Thus they correspond, in thermodynamic terms, not to microstates but to macrostates, and competing fixed points correspond to coexisting phases of the system.  The accurate fixed point corresponds to a ferromagnetic phase which is correlated with the true group assignment, and the trivial fixed point corresponds to a paramagnetic phase.  The condensation transition is the point at which the free energy branches corresponding to these two phases cross, making one phase thermodynamically favored over the other at equilibrium.

The words ``at equilibrium'' are crucial here, implying that we have the luxury of sampling the entire state space of group assignments.  Since the state space is of exponential size as a function of $n$, this is typically not possible for a polynomial-time algorithm.  Thus even if the accurate fixed point has a higher likelihood it may be difficult to find it.  The situation is analogous to that of a glassy material: the lowest free energy of such a system may be attained in the crystalline state, but if that state is surrounded by a high free energy barrier---corresponding dynamically to having an exponentially small basin of attraction---then at reasonable timescales we will remain in the trivial paramagnetic state, and fail to find the true equilibrium.  The hard/easy transition in community detection is the point at which the free energy barrier disappears, so that it becomes dynamically easy for the system to reach the ferromagnetic state and accurately detect the community structure.

We can carry the physical analogy of a first-order transition further.  Suppose we start in the accurate (ferromagnetic) phase, just above the hard/easy transition, and then slowly decrease $\cin-\cout$ so that we enter the coexistence region.  We do this in ``adiabatic'' fashion, making only incremental changes to the structure of the network---adding, removing, or moving edges one by one---iterating the belief propagation equations to convergence after each change, starting from the previous fixed point.  The net result will be that we stay at the accurate fixed point even as we pass the easy/hard transition and enter the regime where that fixed point would be \emph{a priori} hard to find.  We will continue to follow the accurate point until it disappears and we shift to the trivial (paramagnetic) phase.  At that point, we can if we wish start increasing $\cin-\cout$ again, rising back through the coexistence region but now staying at the trivial fixed point until we once again pass the hard/easy transition, where the trivial fixed point destabilizes and we jump back to the accurate one, corresponding to spontaneous magnetization.  In this way we can trace out a hysteresis loop in the behavior of the system; the transitions at which the trivial and nontrivial fixed points become unstable or disappear corresponding to the spinodal lines at the boundaries of the loop.

While it is true that belief propagation rarely finds the accurate fixed point in the coexistence region when the beliefs are randomly initialized, it is still possible to find it if we initialize the beliefs in the right way.  In Section~\ref{sec:numerics} we show the results of numerical calculations where the beliefs are initialized at the known true community assignments of the nodes~$s_i$, meaning we set $\mu_a^{i\to j} = \delta_{a,s_i}$, where $\delta_{a,b}$ is the Kronecker delta.  This initialization places the beliefs sufficiently close to the accurate fixed point that the process reliably converges to it.  In real-world applications one does not know the true assignments of the nodes to communities so this calculation is not possible---finding those assignments is the entire point of performing belief propagation in the first place---but we may still be able to find the accurate fixed point if we have some side information or ``metadata'' about the group assignment that allows us to guess sufficiently good initial values of the beliefs.  This is roughly what happens in the semisupervised case mentioned in the introduction, where we are given the correct labels of a fraction of the nodes.  This kind of information lowers the hard/easy transition, allowing us to find the accurate fixed point at lower values of $\cin-\cout$~\cite{Zhang2014phase,Steeg2013}.  In a similar way, we will see that making the groups unequal lowers the hard/easy transition and shrinks the coexistence region, until, at a critical amount of asymmetry, it removes the detectability transition altogether.

\section{Networks with unequal groups}
\label{sec:unequal}

The purpose of this paper is to understand how the detectability results reviewed in the previous section change when the community structure is asymmetric, i.e.,~when we go from equally sized groups to unequal ones.  In fact, the key question is not whether the groups have unequal sizes, but rather whether they have unequal degrees.  If they do then the trivial fixed point $\mu_a^i=\gamma_a$ no longer exists, and we can no longer identify the hard/easy transition with a simple linear stability analysis.

Here we explore two complementary approaches to this problem.  In the first, we approximate the fixed point by a series expansion about the limit of weak structure.  In the second we approximate it by performing only a finite number of iterations of the belief propagation equations.

\subsection{Series expansion}

In our first approach, we expand the equations for the case of unequal groups about the weak-structure limit, i.e.,~about the limit where $\cin=\cout$.  That is, we choose unequal sizes~$\gamma_a$ for the groups then expand in powers of the strength $\cin-\cout$ of the community structure.  This also results in different average degrees for the groups (which, as we have said, is really the crucial point): from Eq.~\eqref{eq:avgc0}, the average degree~$c_a$ of a node in group~$a$ is
\begin{equation}
c_a = \sum_b c_{ab} \gamma_b = \cout + (\cin-\cout) \gamma_a,
\label{eq:defca}
\end{equation}
so that nodes in larger groups (larger~$\gamma_a$) have higher degree on average whenever $\cin > \cout$.  Thus we can use the node degrees as a guide to community membership.  As we will see, the belief propagation equations employ this local degree information to estimate communities with success better than a random guess, and moreover they spread that information to neighboring nodes to improve the results still further.  The calculation is as follows.

First, note that the average degree in the network as a whole is
\begin{equation}
c = \sum_a \gamma_a c_a = \cout + (\cin-\cout) \bar\gamma,
\label{eq:avgc}
\end{equation}
where
\begin{equation}
\label{eq:bargamma}
\bar\gamma = \sum_a \gamma_a^2
\end{equation}
is the expected size of the community to which a randomly chosen node belongs.  Equivalently, $\bar\gamma$ is the fraction of nodes we would assign to the correct communities purely by chance if we were to place the correct number $n\gamma_a$ of nodes randomly in each group~$a$.

We now expand around the case $\cin=\cout$ by fixing the mean degree~$c$, Eq.~\eqref{eq:avgc}, and varying the difference
\begin{equation}
\epsilon = \cin - \cout .
\label{eq:epsilon}
\end{equation}
This fixes the values of $\cin$ and~$\cout$ uniquely to be $\cin = c + (1-\bar\gamma) \epsilon$ and $\cout = c - \bar\gamma \epsilon$ or equivalently we can write
\begin{equation}
c_{ab} = c + (\delta_{ab} - \bar\gamma) \epsilon.
\label{eq:cdelta2}
\end{equation}

In the limit $\epsilon \to 0$, where $\cin=\cout$, there is no correlation between the community structure and the topology of the network.  Thus the network data tell us nothing and the probability of a node belonging to any group~$a$ is simply equal to the prior probability~$\gamma_a$.  Indeed, it is easy to check in this case that the sole solution of the belief propagation equations is $\mu_a^{i\to j} = \gamma_a$.

We expand about this point in powers of $\epsilon$ thus:
\begin{equation}
\mu_a^{i\to j} = \gamma_a \bigl( 1 + \alpha_a^{i\to j} \epsilon
                  + \ldots \bigr)
\label{eq:mudelta}
\end{equation}
for some coefficients $\alpha_a^{i\to j}$ and expand the marginals similarly:
\begin{equation}
\mu_a^{i} = \gamma_a \bigl( 1 + \alpha_a^{i} \epsilon
                  + \ldots \bigr).
\label{eq:mudelta-marg}
\end{equation}
Since $\sum_a \mu_a^{i\to j} = \sum_a \mu_a^i = 1$ and $\sum_a \gamma_a = 1$, we have
\begin{equation}
\sum_a \gamma_a \alpha_a^{i\to j} 
= \sum_a \gamma_a \alpha_a^i 
= 0.
\label{eq:sumrulea}
\end{equation}
Substituting Eq.~\eqref{eq:mudelta} into Eq.~\eqref{eq:bp} and keeping
terms to first order in~$\epsilon$, we get
\begin{align}
\mu_a^{i\to j} = \frac{\gamma_a}{Z_{i\to j}} 
&\exp\biggl( -\frac{1}{n} \sum_k \sum_b c_{ab} \gamma_b \bigl( 1 + \alpha_b^k \epsilon \bigr) \biggr) \nonumber \\
&\times \prod_{k\in\partial i\backslash j}
      \sum_b c_{ab} \gamma_b \bigl( 1 + \alpha_b^{k\to i} \epsilon \bigr). 
\label{eq:series1}
\end{align}
The sum in the exponential is 
\begin{align*}
\sum_b c_{ab} &\gamma_b \bigl( 1 + \alpha_b^k \epsilon \bigr)
= c_a + \epsilon \sum_b c_{ab} \gamma_b \alpha_b^k \\
&= c_a + \epsilon \Bigl[ \cout \sum_b \gamma_b \alpha_b^k + (\cin-\cout) \gamma_a \alpha_a^k \Bigr] \\
&= c_a + \epsilon^2 \gamma_a \alpha_a^k 
= c_a + O(\epsilon^2) ,
\end{align*}
where we have used Eqs.~\eqref{eq:avgc0}, \eqref{eq:epsilon}, and~\eqref{eq:sumrulea}.  Similarly,
\begin{align*}
\sum_b c_{ab} \gamma_b ( 1 + \alpha_b^{k\to i} \epsilon ) 
= c_a + \epsilon^2 \gamma_a \alpha_a^{k\to i} 
= c_a + O(\epsilon^2) .
\end{align*}
Using these expressions in~\eqref{eq:series1}, along with Eq.~\eqref{eq:avgc} again, we get
\begin{equation}
\mu_a^{i\to j} = \frac{\gamma_a}{Z_{i\to j}} \,\e^{-c_a} c_a^{d_i-1},
\label{eq:museries}
\end{equation}
where $Z_{i\to j}$ is the appropriate normalizing constant as usual.  Notice that $\mu_a^{i\to j}$ is independent of~$j$ at this order, meaning that a vertex sends the same message to each of its neighbors.

Similarly, we can calculate the one-node marginal probabilities~$\mu^i_a$ from Eq.~\eqref{eq:bpmu} and we get
\begin{align}
\mu_a^i &= \frac{\gamma_a}{Z_i} \,\e^{-c_a} c_a^{d_i}.
\label{eq:series2}
\end{align}
This tells us that nodes with higher degree~$d_i$ will have a higher probability of being placed in groups where the average degree~$c_a$ is higher, while those with lower degree will have a higher probability of being placed in groups with lower average degree.  In other words, the algorithm will divide the nodes according to their degrees.  As a result, whenever $\epsilon > 0$ there is no regime in which we do no better than chance.

Specifically, since nodes in group $a$ have degrees which are Poisson-distributed with mean $c_a$, Eq.~\eqref{eq:series2} implies that the marginals are exactly equal to the posterior probabilities of the groups given the degree, since
\begin{align}
\Pr[s_i=a \mid d_i]
  &= {\Pr[s_i=a]\over \Pr[d_i]} \Pr[d_i \mid s_i=a] \nonumber\\
  &= \frac{\gamma_a}{Z_i} \,\e^{-c_a} c_a^{d_i} = \mu_a^i , 
\label{eq:bayes-degree}
\end{align}
where $\gamma_a = \Pr[s_i=a]$ by definition and $Z_i = d_i! \Pr[d_i]$ is the required normalization constant.  This is the Bayes-optimal conclusion that we can reach about $i$'s group membership, given no information except its degree, or equivalently given only its radius-$1$ neighborhood in the network.

That only the radius-$1$ neighborhood enters into this calculation is a result of the fact that, in the weak-structure limit where we treat $\epsilon$ to first order, belief propagation transmits information only one step along the edges of the network before it reaches a fixed point.  If we calculate the next order in the series, treating terms up to second order in $\epsilon$, we will find ourselves taking the radius-$2$ neighborhood into account, classifying nodes based on their own degree and the degrees of their neighbors, and so on.  This suggests an alternative approach which we describe in the following section.

\subsection{Finite iteration of the\\belief propagation equations}
\label{sec:bpfinite}

As discussed above, a series expansion of the belief propagation equations produces a set of approximations for the fixed point that depend on information from a neighborhood of increasing radius around the node of interest.  This prompts us to consider an alternative approach in which we look at the behavior of the belief propagation algorithm after a finite number of iterations of the update equations~\eqref{eq:bp}.  Since each iteration corresponds to each node passing its current information to its neighbors, $t$~iterations mean that each node receives information from its neighbors out to distance~$t$.

Suppose we start with messages derived from nothing but the prior on group assignments, 
i.e.,~$\mu^{i\to j}_a = \gamma_a$ for all $i,j,a$, 
and apply belief propagation for a single step. 
After one iteration of Eq.~\eqref{eq:bp} the new values of the beliefs will be
\begin{equation}
\mu_a^{i\to j}(1) = \frac{\gamma_a}{Z_{i\to j}(1)} \,\e^{-c_a} c_a^{d_i-1},
\end{equation}
where $Z_{i\to j}(1)$ is the appropriate normalizing constant as usual and we have made use of Eq.~\eqref{eq:avgc0}.  These values are identical to those derived from the first-order expansion of the previous section, Eq.~\eqref{eq:museries}.  Similarly, from Eq.~\eqref{eq:bpmu}, the one-node marginal probabilities are
\begin{equation}
\mu_a^i(1) = {1\over Z_i(1)} \,\gamma_a \e^{-c_a} c_a^{d_i} 
  = \Pr[s_i = a \mid d_i],
\label{eq:step1}
\end{equation}
the same again as in the previous section, Eq.~\eqref{eq:series2}.  And, as previously, this is the optimal Bayesian classification of the nodes based on their radius-1 neighborhoods in the network: that is, based only on how many neighbors they have, but without any further information about those neighbors---see Eq.~\eqref{eq:bayes-degree}.

If we perform a second step of belief propagation, we get
\begin{equation}
\mu_a^{i\to j}(2) = {\gamma_a \e^{-c_a}\over Z_{i\to j}(2)}
  \prod_{k\in\partial i\backslash j} {1\over Z_{k\to i}(1)} 
  \sum_b \gamma_b c_{ab} 
  \e^{-c_b} c_b^{d_k-1}
\end{equation}
and
\begin{equation}
\label{eq:step2}
\mu_a^i(2) = {\gamma_a \e^{-c_a}\over Z_i(2)}
  \prod_{k\in\partial i} {1\over Z_{k\to i}(1)} 
  \sum_b \gamma_b c_{ab} 
  \e^{-c_b} c_b^{d_k-1}.
\end{equation}
Now the marginals depend both on $i$'s degree and the degrees of its neighbors, i.e., on $i$'s neighborhood of radius $2$.  And again this is the optimal Bayesian classification given this information and no other, as we can see by noting that if $k$ is a neighbor of $i$ and is of type $b$ then its so-called \defn{excess degree}---that is, the number of neighbors $k$ has in addition to $i$---is Poisson-distributed with mean~$c_b$.  Thus
\begin{equation}
  \Pr[d_k \mid k \in \partial i, s_k=b] = \frac{\e^{-c_b} c_b^{d_k-1}}{(d_k-1)!} \, . 
\end{equation}
Furthermore, the definition of the block model gives
\begin{equation}
\Pr[ k \in \partial i \mid s_k=b, s_i=a] 
= p_{ab} \, ,
\end{equation}
and so
\begin{equation}
\Pr[ s_k=b \mid k \in \partial i, s_i=a] 
= \frac{\gamma_b p_{ab}}{\sum_{b'} \gamma_{b'} p_{ab'}}
= \frac{\gamma_b c_{ab}}{c_a} \, . 
\end{equation}
Now, applying Bayes' rule and summing over all possible types of $i$'s neighbors (which are unknown to us) gives the following probability that $i$ is of type $a$, given $i$'s degree and those of its neighbors:
\begin{align}
\!\!\Pr[ s_i = a &\mid d_i , \{ d_k \} ] \propto
\gamma_a \Pr[d_i, \{ d_k \} \mid s_i=a] 
  \nonumber \\
&= \gamma_a \Pr[d_i \mid s_i=a] \,\prod_{k \in \partial i} \Pr[d_k \mid k \in \partial i, s_i=a]
  \nonumber \\ 
&= \gamma_a \frac{\e^{-c_a} c_a^{d_i}}{d_i!} 
\prod_{k \in \partial i} \sum_b \Pr[ s_k=b \mid k \in \partial i, s_i=a] \nonumber \\
& \hspace{1in}{} \times \Pr[d_k \mid k \in \partial i, s_k=b]
  \nonumber \\
&= \gamma_a \frac{\e^{-c_a} c_a^{d_i}}{d_i!} 
\prod_{k \in \partial i} \sum_b \frac{\gamma_b c_{ab}}{c_a} \,\frac{\e^{-c_b} c_b^{d_k-1}}{(d_k-1)!} 
  \nonumber \\
&\propto \gamma_a \e^{-c_a}
\prod_{k \in \partial i} \frac{1}{(d_k-1)!} \sum_b \gamma_b c_{ab} \e^{-c_b} c_b^{d_k-1},
\end{align}
which (after normalization) matches Eq.~\eqref{eq:step2}.

These results extend naturally to any number~$t$ of iterations: if we start with uniform messages and iterate belief propagation $t$ times we get the Bayes-optimal estimate of $i$'s marginals based on its network neighborhood of radius~$t$.  Indeed, the belief propagation equations are equivalent simply to applying Bayes' rule locally, updating $i$'s marginal based on those of its neighbors with the assumption that $i$'s neighbors are independent of each other.  This holds exactly on trees and, therefore, also on locally tree-like networks such as those generated by the stochastic block model, on the radius-$t$ neighborhood of almost all vertices.  Thus, iterating belief propagation $t$ times is an asymptotically optimal algorithm for labeling nodes of a stochastic block model network based on local information up to $t$ steps away in the network.

Since we know that the local neighborhood carries information about group membership in the case of asymmetric groups, this allows us to conclude that belief propagation, starting from messages equal to the prior probabilities, will always label the nodes better than a random guess.  It is by no means guaranteed, however, that a local calculation of this kind must give the best possible answer.  It is possible that some nonlocal calculation could do better and indeed this is exactly what happens in the coexistence region for the case $q>4$.  In this region the local calculation does do better than a random guess, but there exists another fixed point that does better still.  Finding this fixed point, however, requires us to start belief propagation very close to it, meaning we have to give the algorithm fundamentally nonlocal information, simultaneously choosing the correct values of the beliefs out to arbitrary distances.

\section{Numerical results}
\label{sec:numerics}

The results of Section~\ref{sec:unequal} suggest that it should be possible to classify nodes into the correct groups with a success rate better than chance for all networks with $\cin>\cout$ when group sizes are unequal or, more generally, when average degrees are unequal.  In this section, we test this prediction with numerical experiments on networks generated by the stochastic block model.  As we will see, our expectations are borne out by the simulations and a number of other phenomena are revealed as well, particularly concerning the picture for networks with larger numbers of communities.  As described in Section~\ref{sec:detectability}, when $q > 4$ there are, for certain parameter regions, two stable fixed points.  When the size or average degrees of the groups are equal, the values of the messages at one of these fixed points (the ``trivial'' fixed point) give no information about community memberships while those at the other give a group assignment strongly correlated with the true one, and there is a first-order phase transition between the two.  When the group sizes are unequal or when the groups have different average degrees a random guess according to the prior probabilities~$\gamma_a$ achieves an accuracy of~$\bar{\gamma}$, Eq.~\eqref{eq:bargamma}, but the calculations of the previous section suggest that even the less good of the two fixed points achieves an accuracy significantly better than this.  Thus in this regime we expect to see a (first-order) phase transition between ``good'' and ``better'' performance, but no regime in which the algorithm fails altogether.

In order to measure the effect of unequal group sizes and degrees, we explore a two-parameter space of block model networks.  The first parameter is the difference $\epsilon=\cin-\cout$ between the densities of in-group and between-group connections, as defined previously in Eq.~\eqref{eq:epsilon}.  The second parameter, which we denote~$\delta$, measures the amount of asymmetry in the groups, i.e.,~how far we are from having equally sized groups.  We define the group sizes~$\gamma_a$ to be
\begin{equation}
\gamma_a = {1\over q} ( 1 + \delta\zeta_a ),
\label{eq:delta}
\end{equation}
where the quantities~$\zeta_a$ are of order~1 and sum to zero, $\sum_a
\zeta_a = 0$.  This choice satisfies the normalization constraint $\sum_a
\gamma_a = 1$ and allows us to go from equal-sized groups at $\delta=0$ to
unequal ones for $\delta>0$.  For the particular simulations performed
here, we consider equally spaced group sizes with
\begin{equation}
\zeta_a = a - \half(q+1) .
\label{eq:zeta}
\end{equation}
For $q=3$, for example, we would have groups of size $\tfrac13$ and $(1 \pm \delta)/3$.  Varying $\delta$ also varies the average group degrees.  From Eq.~\eqref{eq:defca} we have
\begin{equation}
\label{eq:ca-delta}
c_a = \cout + {\epsilon\over q} ( 1 + \delta\zeta_a ),
\end{equation}
so the groups have different average degrees whenever $\delta > 0$.

To quantify our success (or lack of success) at identifying the planted community structure, we calculate the \defn{overlap} between the planted and detected communities, equal to the fraction of nodes assigned to their correct communities by the algorithm.  There is, however, some ambiguity about how the overlap is defined, given that belief propagation does not uniquely assign nodes to single communities but rather gives us the marginal probabilities~$\mu_a^i$ with which the nodes belong to each community.  Conventionally, one removes this ambiguity by assigning each node to the community it has the highest probability of belonging to.  Then the overlap~is
\begin{equation}
Q = {1\over n} \sum_i \delta(s_i,\argmax_a \mu_a^i),
\label{eq:ovl}
\end{equation}
where $s_i$ is the planted community of node~$i$ as previously,
$\delta(i,j)$ is the Kronecker delta, and $\argmax_a f(a)$ denotes the
value of $a$ that maximizes~$f(a)$.

This measure has some problems, however.  It throws away a lot of information contained in the marginals when a node has a significant probability of belonging to more than one group.  Moreover, it can assign a node to a group even if the probability it belongs there is only a little above~$1/q$, so for large~$q$ the most probable assignment may be quite unlikely to be correct.  An alternative measure that takes these issues into account is the \defn{marginal overlap}
\begin{equation}
Q_\mu = {1\over n} \sum_i \mu_{s_i}^i,
\label{eq:movl}
\end{equation}
which is equal to the total fraction of nodes that would be assigned to the correct communities if communities were assigned randomly in proportion to their marginal probabilities.

Note that these two definitions of the overlap have different values in the weak-structure limit where the marginal probabilities are equal to the group sizes $\mu_a^i = \gamma_a$.  In the case where each node is assigned to its most likely group we end up putting all nodes in the largest group in the weak-structure limit, which means that the fraction of correctly assigned nodes is
\begin{equation}
Q = \max_a \gamma_a = {1\over q} \bigl[ 1 + \half(q-1)\delta \bigr]
\label{eq:ovlweak}
\end{equation}
for the choice of group sizes in Eq.~\eqref{eq:zeta}.  In contrast, the
value of the marginal overlap in the weak-structure limit~is
\begin{equation}
Q_\mu = {1\over n} \sum_i \gamma_{s_i} = \bar{\gamma}
    = {1\over q} \bigl[ 1 + \tfrac{1}{12} (q^2-1) \delta^2 \bigr].
\end{equation}

\subsection{Performance of belief propagation}

\begin{figure*}
\centering
\includegraphics[width=0.45\textwidth]{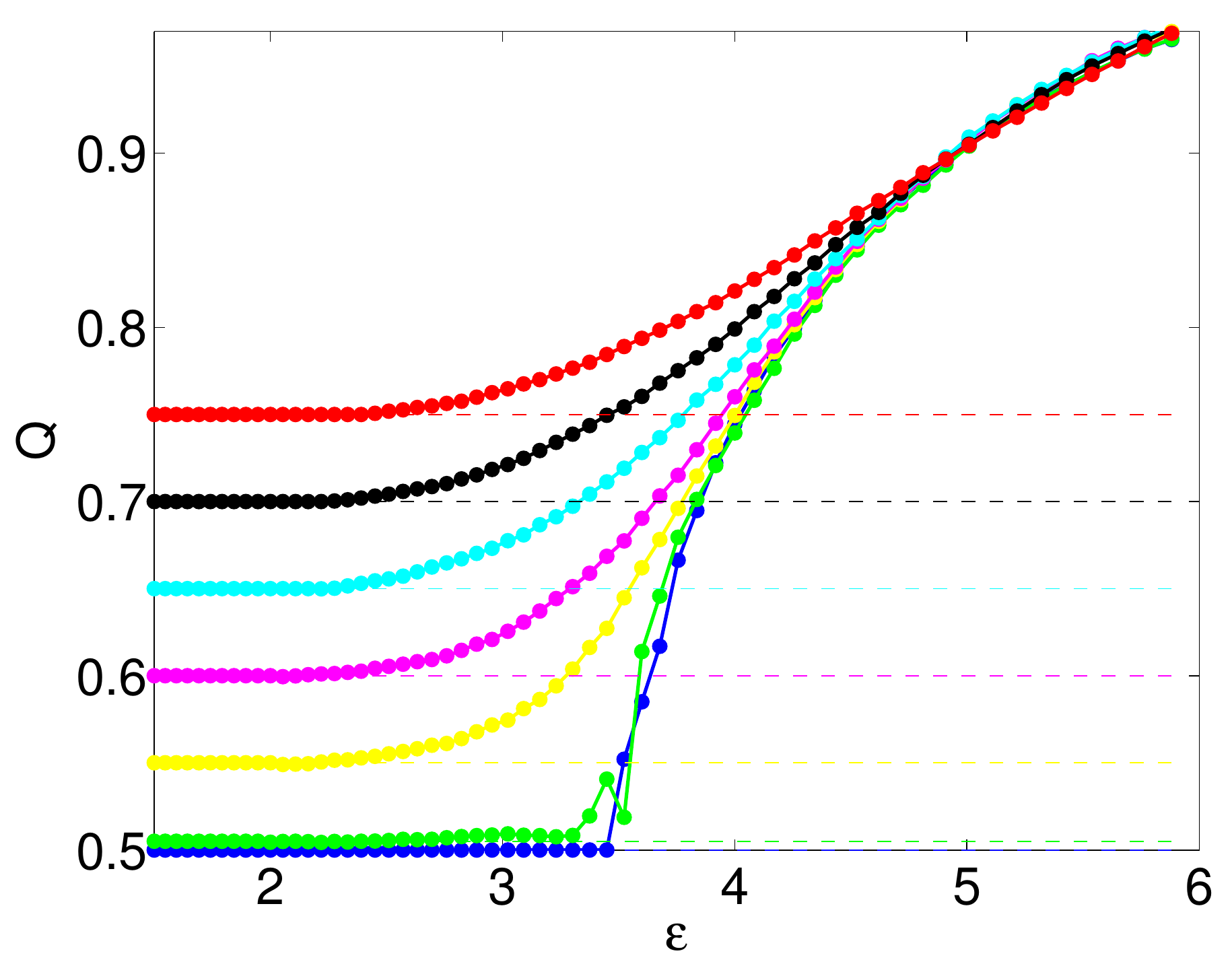}\qquad
\includegraphics[width=0.45\textwidth]{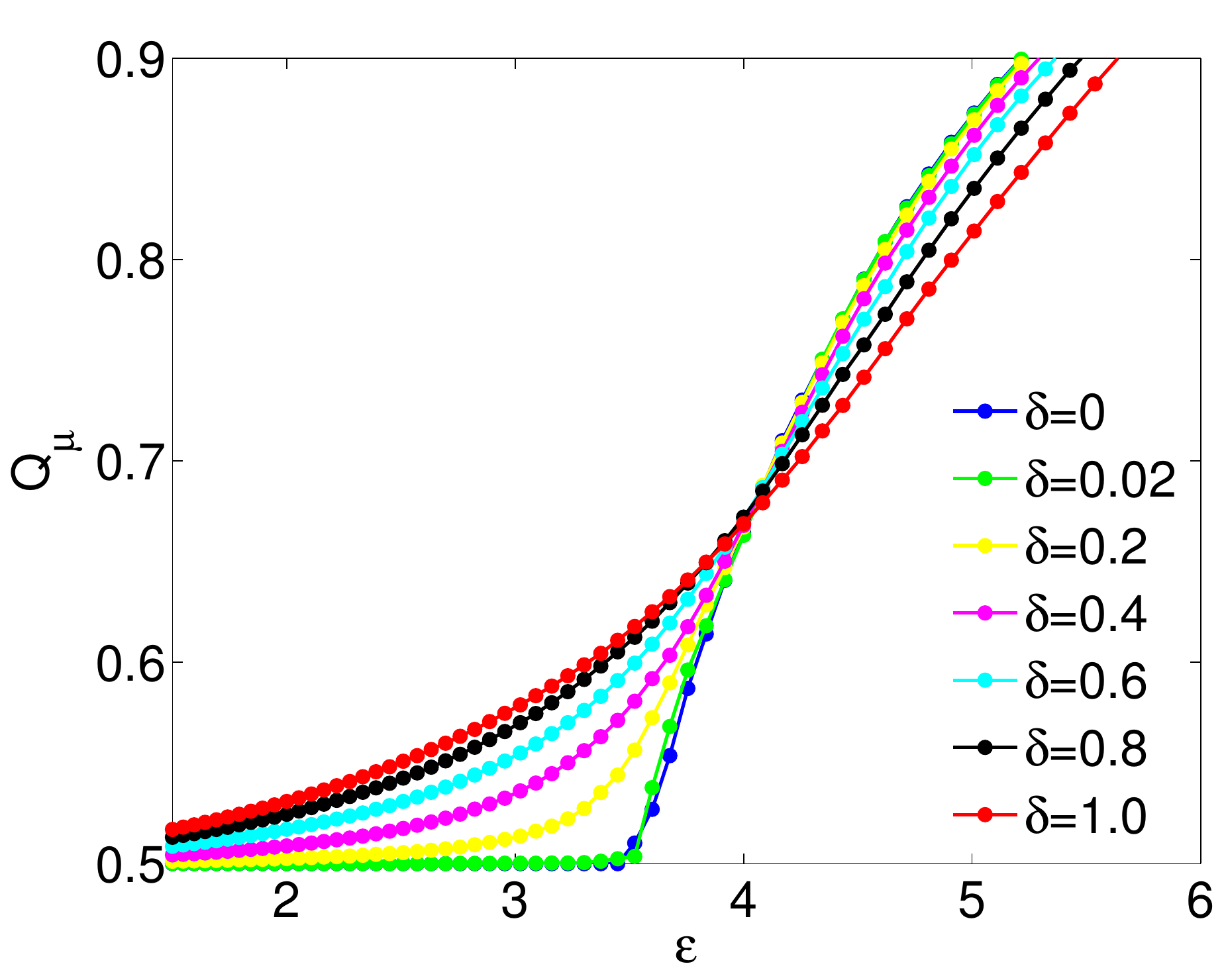} 
\caption{The overlap $Q$, Eq.~\eqref{eq:ovl}, and the marginal overlap $Q_\mu$, Eq.~\eqref{eq:movl}, for belief propagation on networks generated by the stochastic block model with $q=2$ groups, $n=10^5$ nodes, average degree $c=3$, and group sizes as given in Eqs.~\eqref{eq:delta} and~\eqref{eq:zeta} a function of $\epsilon=\cin-\cout$ for various values of $\delta$.  Increasing $\delta$ corresponds to greater differences between the group sizes and average degrees.  The dashed lines in the left panel are the expected values in the weak-structure (i.e.,~$\epsilon=0$) limit, Eq.~\eqref{eq:ovlweak}.  Note how the sharp detectability transition disappears for $\delta > 0$; both overlaps are smooth functions of the block model parameters.}
\label{fig:q2ovl}
\end{figure*}

 \begin{figure*}
 \centering
 \includegraphics[width=0.51\textwidth]{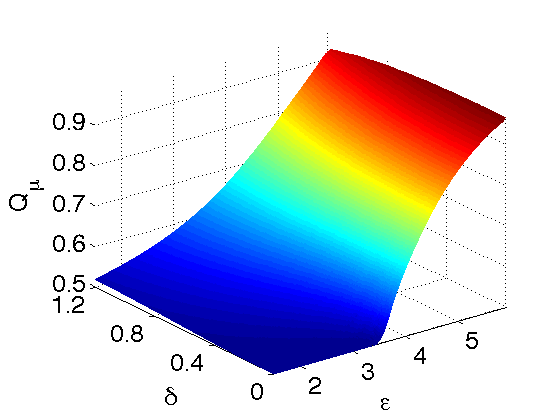}
 \includegraphics[width=0.48\textwidth]{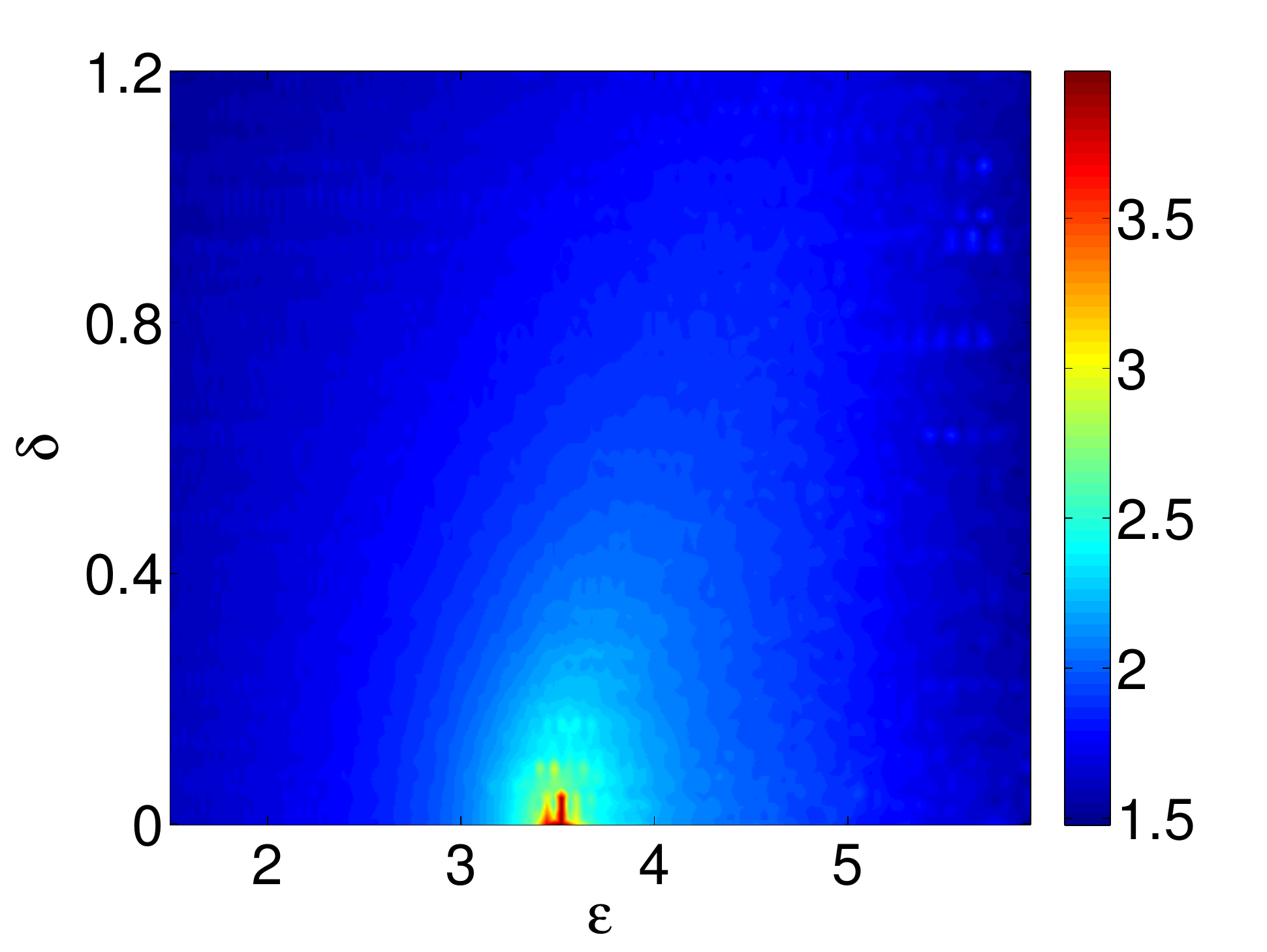} 
 \caption{The marginal overlap $Q_\mu$ (left) and log (base~10) of the convergence time (right) as a function of $\epsilon$ and $\delta$ for networks with $q=2$ groups, size $n=10^5$, and average degree $c=3$.  The overlap is a smooth function except at the detectability transition for equal-sized groups, which occurs when $\delta=0$ and $\epsilon = 2\sqrt{c}$.  This is also the only place where the convergence time diverges.  Thus for $q=2$ and $\delta > 0$ there is no detectability transition.}
 \label{fig:q2i1}
 \end{figure*}

Figure~\ref{fig:q2ovl} shows the overlap (left) and marginal overlap (right) for networks with $q=2$ groups.  For these calculations we generated networks with $n=100\,000$ nodes, average degree $c=3$, and various values of the parameters~$\delta$ and~$\epsilon$, then ran the belief propagation algorithm starting from random initial messages.

For the case~$\delta=0$, where the two communities are of equal size and equal average degree, we see that there is, as in~\cite{DKMZ11a,DKMZ11b}, a phase transition at $\epsilon_c = 2 \sqrt{c} = 3.46\ldots$ (see Eq.~\ref{eq:threshold}) from a regime where the overlap is~$\half$ by either definition---no better than a random guess---to one with overlap strictly greater than~$\half$.  For $\delta>0$, however, where the communities have unequal size and unequal average degree, we see that the algorithm does better than chance whenever $\epsilon > 0$; moreover, the detectability transition disappears, i.e.,~the overlap is a smooth function of~$\epsilon$.  Figure~\ref{fig:q2i1} provides an alternative visualization of the behavior of the system.  Here we show the overlap~$Q$ (left) and the convergence time (right) in the $\epsilon$-$\delta$ plane.  These figures make the lack of a sharp detectability transition particularly clear: the only place where $Q$ is not a smooth function, and the only place where the convergence time diverges, is at the equal-group detectability transition, when $\delta=0$ and $\epsilon=\epsilon_c$.

\begin{figure*}
\centering
\includegraphics[width=0.45\textwidth]{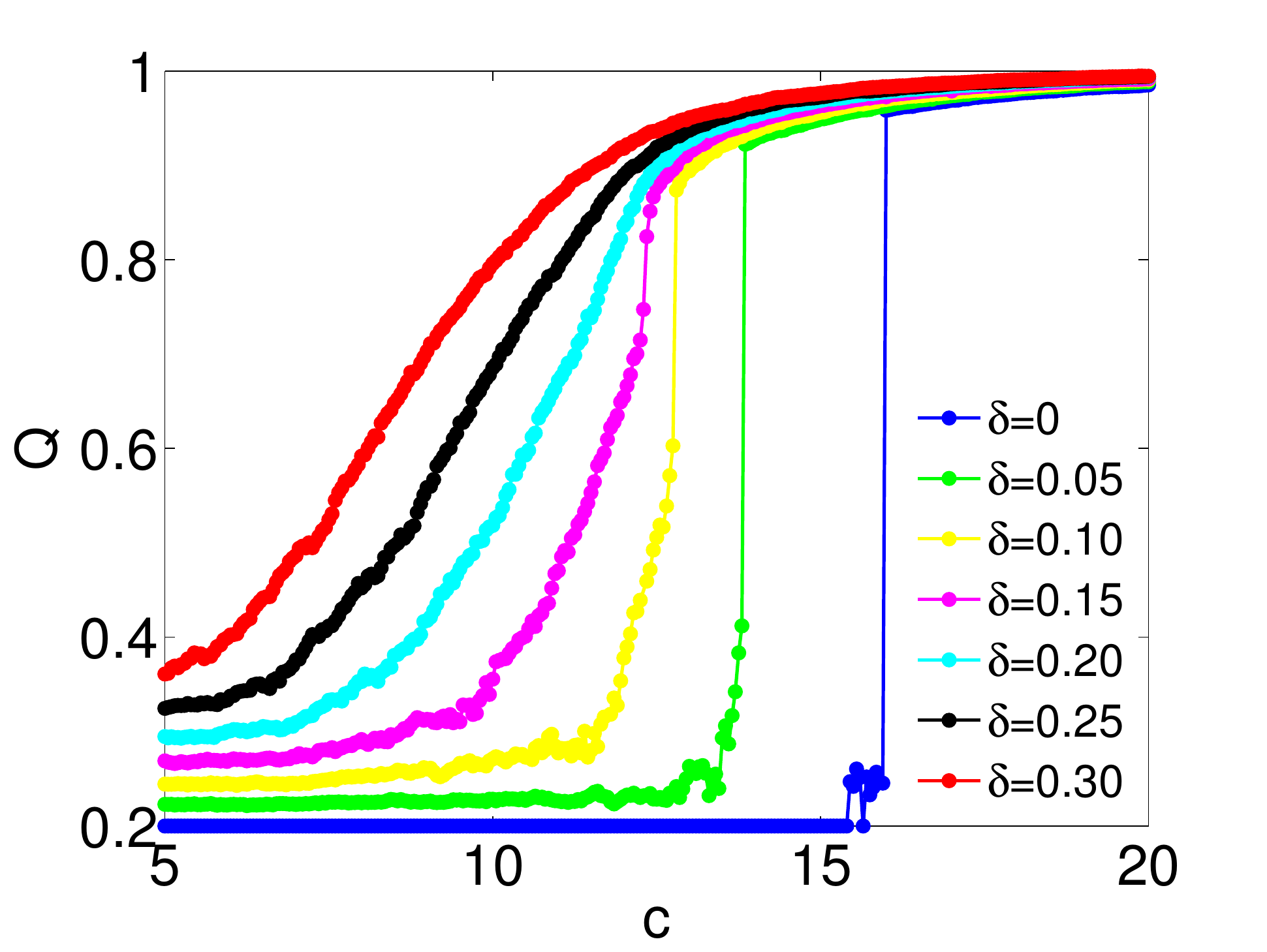}\qquad
\includegraphics[width=0.45\textwidth]{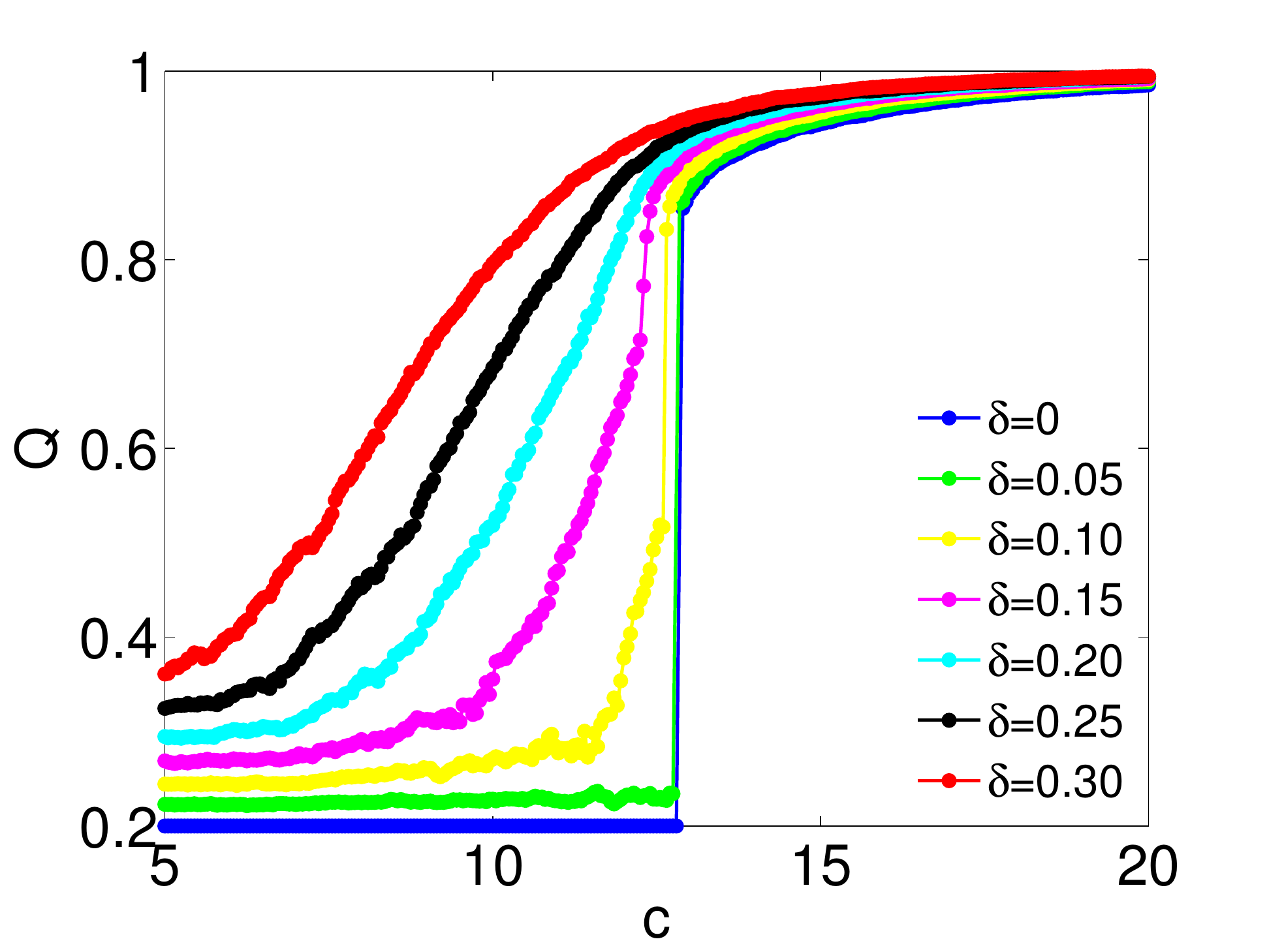}
\caption{The overlap $Q$ for belief propagation on fully disassortative networks generated by the stochastic block model with $q=5$ groups, $n=10^5$ nodes, and various values of~$\delta$ as indicated, as a function of average degree~$c$.  In the left panel we initialize the beliefs with uniform random values; in the right panel we initialize them with the true (planted) communities.  For small values of~$\delta$ there is a range of $c$ where the latter initialization gives a higher overlap, indicating a second and better fixed point with a small basin of attraction.}
\label{fig:q5ovl}
\end{figure*}

\begin{figure*}
\centering
\includegraphics[width=0.48\textwidth]{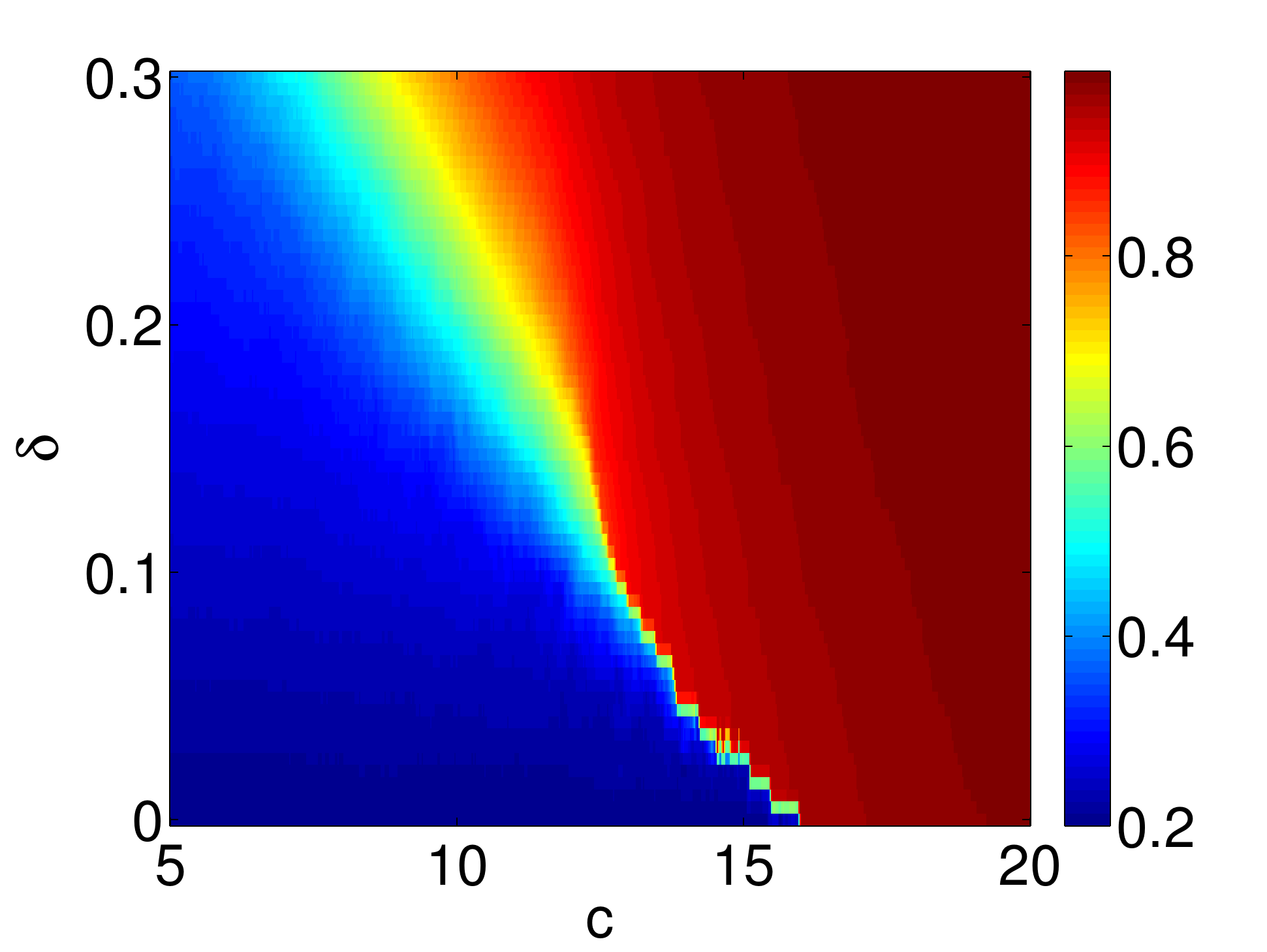} 
\includegraphics[width=0.48\textwidth]{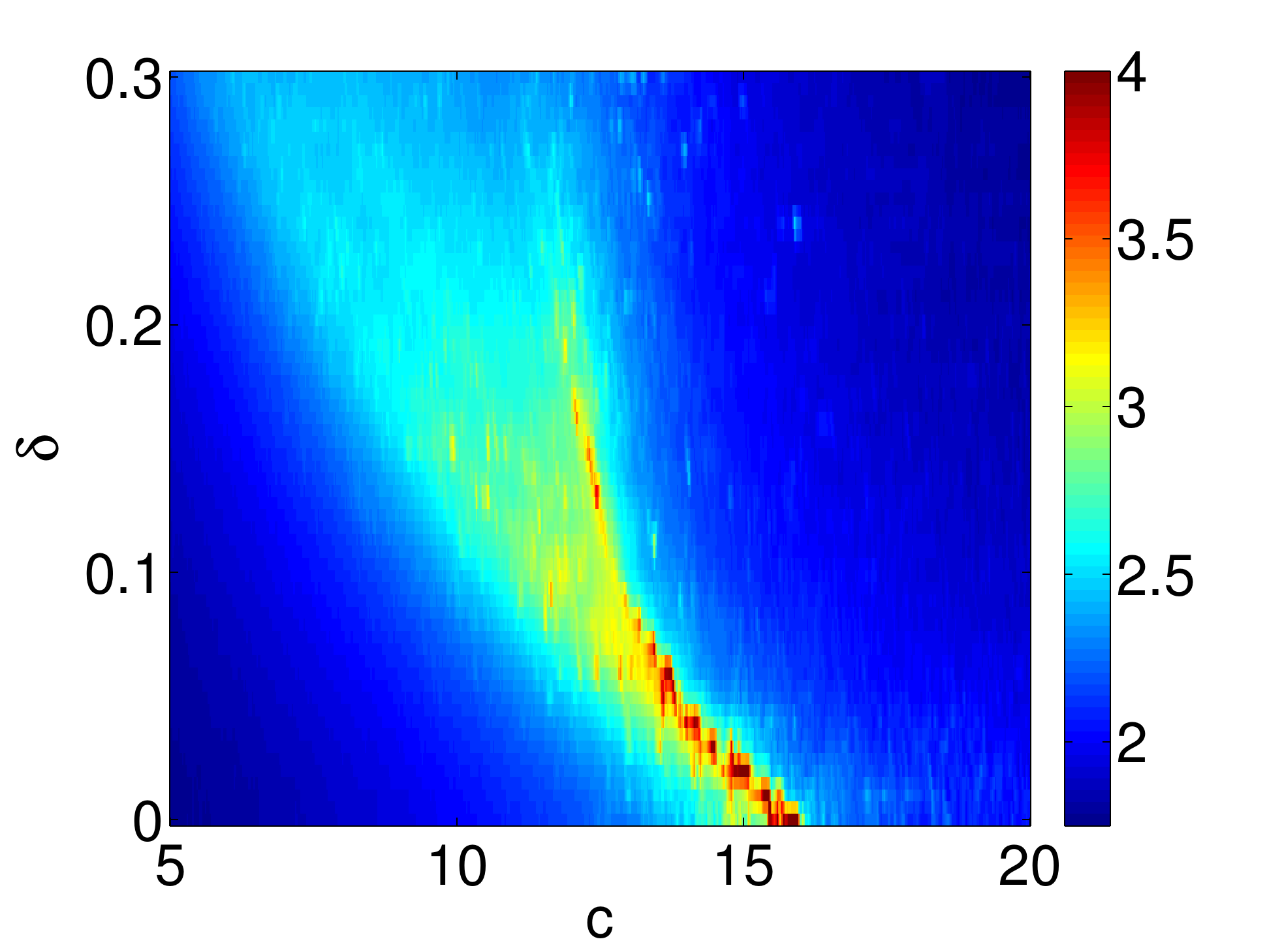} 
\caption{The overlap~$Q$ (left) and log (base 10) of the convergence time (right) as a function of $c$ and $\delta$ for totally disassortative networks with $q=5$ groups and $n=10^5$ nodes.  The beliefs are initialized with random values in both panels.  The first-order hard/easy transition is visible as a line in the $c$--$\delta$ plane where the overlap jumps discontinuously and the convergence time diverges.  The height of the discontinuity decreases with increasing~$\delta$ until we reach a critical point at which it vanishes at a second-order transition.  Above this point the overlap is a smooth function of $c$ and $\delta$ and there is no detectability transition.}
\label{fig:color:i1}
\end{figure*}

We have also performed tests on networks with three and four groups and find similar behavior.  For more than four groups we expect qualitatively different behavior as described above---a first-order transition with a coexistence region below the hard/easy transition, characterized by the simultaneous coexistence of two stable fixed points.  Unfortunately, clear numerical confirmation of this behavior is harder to obtain.  The coexistence region is difficult to see because the range of $\epsilon$ it spans is quite narrow for assortative networks.  As observed in Ref.~\cite{DKMZ11b}, however, the behavior is clearer in the disassortative case, and particularly in the fully disassortative case of a network that has connections only between different groups and none within groups.  (Community detection in this case is equivalent to a ``planted graph coloring problem.''  In computer science a \emph{$q$-coloring} of a graph is a coloring or labeling of the vertices with $q$ different labels such that no vertices with the same label have an edge between them.  Our problem is equivalent to one in which we generate a random graph that we know to be colorable in this way by first assigning the labels and then adding edges only between unlike labels.  Then we discard the labels and try to recover them again based only on the structure of the graph.)  Since $\cin=0$ in a totally disassortative network, our parameter~$\epsilon$ is just $-\cout$ in this case while the average degree, Eq.~\eqref{eq:avgc}, is $c = \cout(1-\bar\gamma)$.  Thus there is no need for separate parameters $c$ and~$\epsilon$: fixing the average degree automatically fixes~$\epsilon$.

Recall that both of the fixed points in the coexistence region are expected to give better-than-random classification of the nodes into communities, but one is expected to perform better than the other.  The two points can be considered perturbations of the ``accurate'' and ``trivial'' fixed points of the equal-groups case.  Roughly speaking, the perturbed ``near-trivial'' fixed point corresponds to inference with local information, starting with the prior and applying belief propagation a few times, while the accurate fixed point corresponds to finding a self-consistent solution with global correlations and considerably higher accuracy.  We expect both fixed points to be locally stable, but for the accurate fixed point to have an exponentially smaller basin of attraction than the near-trivial one.

To test this hypothesis we perform two separate sets of experiments.  In the first we initialize belief propagation with uniformly random messages~$\mu_a^{i \to j}$ (up to normalization).  With this random initialization belief propagation typically converges to the near-trivial fixed point, unless we are above the hard/easy transition at which this point becomes unstable.  In the second set of simulations we initialize belief propagation with messages corresponding to the true communities that we planted in the network, $\mu_a^{i\to j} = \delta_{a,s_i}$.  With this planted initialization belief propagation typically converges to the accurate fixed point, unless we are below the spinodal transition at which this point disappears.  Thus, above and below the coexistence region we expect these two sets of experiments to converge to the same solution, while within the coexistence region we expect them to give different solutions, with the random initialization giving a lower overlap than the planted one.

\begin{figure}
\centering
\includegraphics[width=\columnwidth]{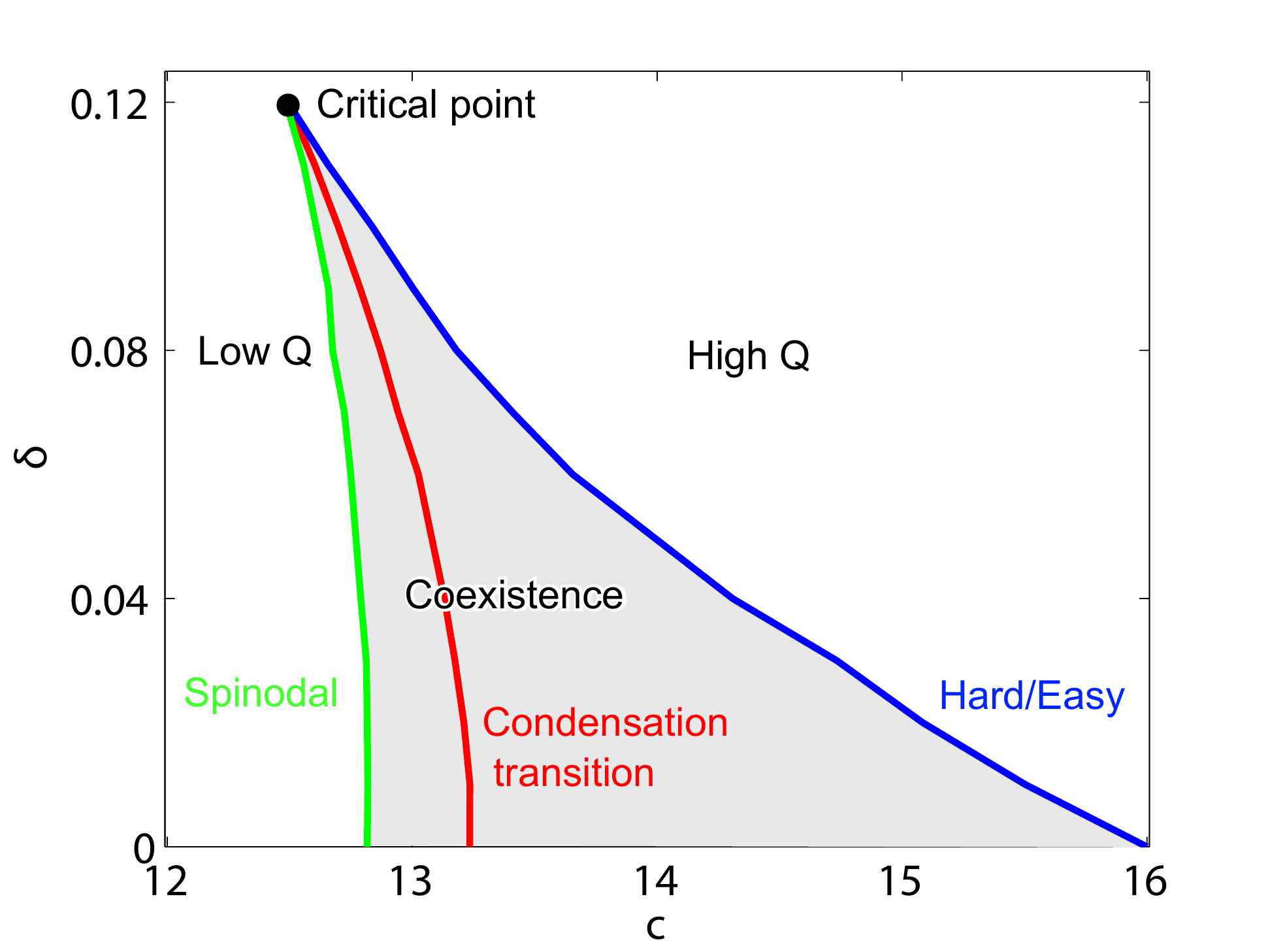} 
\caption{Phase diagram in the $c$--$\delta$ plane for the $q=5$ fully disassortative case described in the text.  The blue curve shows where the near-trivial fixed point becomes unstable (also called the Kestum--Stigum or easy/hard transition); the green curve is the point at which the accurate fixed point disappears.  The gray area between the two is the coexistence region in which both fixed points are stable and belief propagation can converge to either depending on the initial conditions.  The red curve is the condensation transition at which the likelihoods cross over; the black dot is the critical point above which there is no phase transition behavior at all.}
\label{fig:spinodal_static}
\end{figure}

Figure~\ref{fig:q5ovl} shows the overlap $Q$ as a function of~$c$ for fully disassortative networks with $q=5$ and $n=100\,000$, with random initial messages (left) and the planted initialization (right), run on the same set of networks in each case.  As the figure shows, the results are indeed as hypothesized above.  For low and high values of~$c$ the two initializations give the same results, as they do also for sufficiently large values of~$\delta$.  For small values of~$\delta$, however, there is a sizable range of values of $c$ where the overlap achieved by belief propagation with random initial messages is significantly lower than that with the planted initialization, indicating the coexistence of two competing fixed points.  For comparison, the hard/easy transition for fully disassortative networks in the equal-group case~\cite{DKMZ11a} is at $c=(q-1)^2=16$.

Figure~\ref{fig:color:i1} again gives a different view of the results, with the left panel showing the overlap achieved by belief propagation with random initial messages in the $c$-$\delta$ plane.  There is a clear curve visible in this plot where the overlap changes discontinuously as the near-trivial fixed point becomes unstable and belief propagation jumps to the accurate fixed point.  Exactly on this curve, the near-trivial fixed point is marginally stable, causing the convergence time to diverge, as shown in the right panel.  Thus there is a hard/easy transition in this case, even though there was none for $q=2$, and it is a first-order transition.  As the asymmetry increases with~$\delta$, however, the size of the discontinuity shrinks and past a certain point (about $\delta=0.12$) it vanishes altogether.  The critical point where it vanishes is a second-order phase transition and beyond this transition the overlap is a smooth function of the block model parameters.  

This behavior is reminiscent of a first-order transition in a spin system with an external field, where the order parameter shows a discontinuity as a function of temperature but the size of the discontinuity decreases and then vanishes at a critical value of the external field~\cite{Franz1997,Zhang2014phase}.
%(Compare this with the two-dimensional Ising model, where there is a %discontinuity in the magnetization as a function of the external field, which %vanishes at a critical value of the temperature; in both cases there is a %line of discontinuities in the phase diagram which vanish at a critical %point.)
In the present case the ``temperature'' comes from the average degree and/or the strength of the community structure, and the ``external field'' comes simply from the topology of the network.

Figure~\ref{fig:spinodal_static} shows the behavior observed in our experiments as a single phase diagram in the $c$--$\delta$ plane.  The blue curve represents the hard/easy transition at which the near-trivial fixed point becomes unstable; to the right of this curve only the accurate fixed point is stable, so all calculations converge to a high overlap, regardless of whether they are initialized randomly or with the planted communities.  The green curve shows the spinodal line where the accurate fixed point disappears; to the left of this curve both initializations converge to the near-trivial fixed point, yielding a relatively low overlap.  In between lies the coexistence region (gray), which extends up to the critical point at $\delta\simeq0.12$; for $\delta$ larger than this there is no phase transition.  Finally, the red curve is the condensation transition mentioned in Section~\ref{sec:detectability}, the line at which the likelihoods (or equivalently the Bethe free energies) of the two fixed points cross.  To the left of this line the algorithm that finds the fixed point with highest likelihood will choose the near-trivial fixed point over the accurate one, and hence fail to detect the communities no matter how much time is allowed.

Note that, even to the right of the hard/easy transition, there can be locally stable fixed points other than the accurate one: when $\epsilon$ is large enough or $\delta$ is small enough, there are also fixed points corresponding to various permutations of the groups.  These permuted fixed points have lower likelihood than the one corresponding to the planted community structure, but for large $\epsilon$ they can have fairly large basins of attraction, causing belief propagation with random initial messages to fall into them fairly often.  (This is the source of some of the fluctuations visible in Fig.~\ref{fig:q5ovl}.)  Nevertheless, we can find the accurate fixed point in this case by performing a reasonable number of independent runs of belief propagation and choosing the fixed point with the highest likelihood.

\subsection{Belief propagation with a finite number of steps}

In this section we investigate the behavior of belief propagation when run for a finite number of steps, as opposed to iterating it until it converges to a fixed point.  As discussed in Section~\ref{sec:bpfinite}, iterating belief propagation $t$ times makes optimal use of local information up to~$t$ steps away, but ignores information further than that.

\begin{figure*}
\centering
\includegraphics[width=0.43\textwidth]{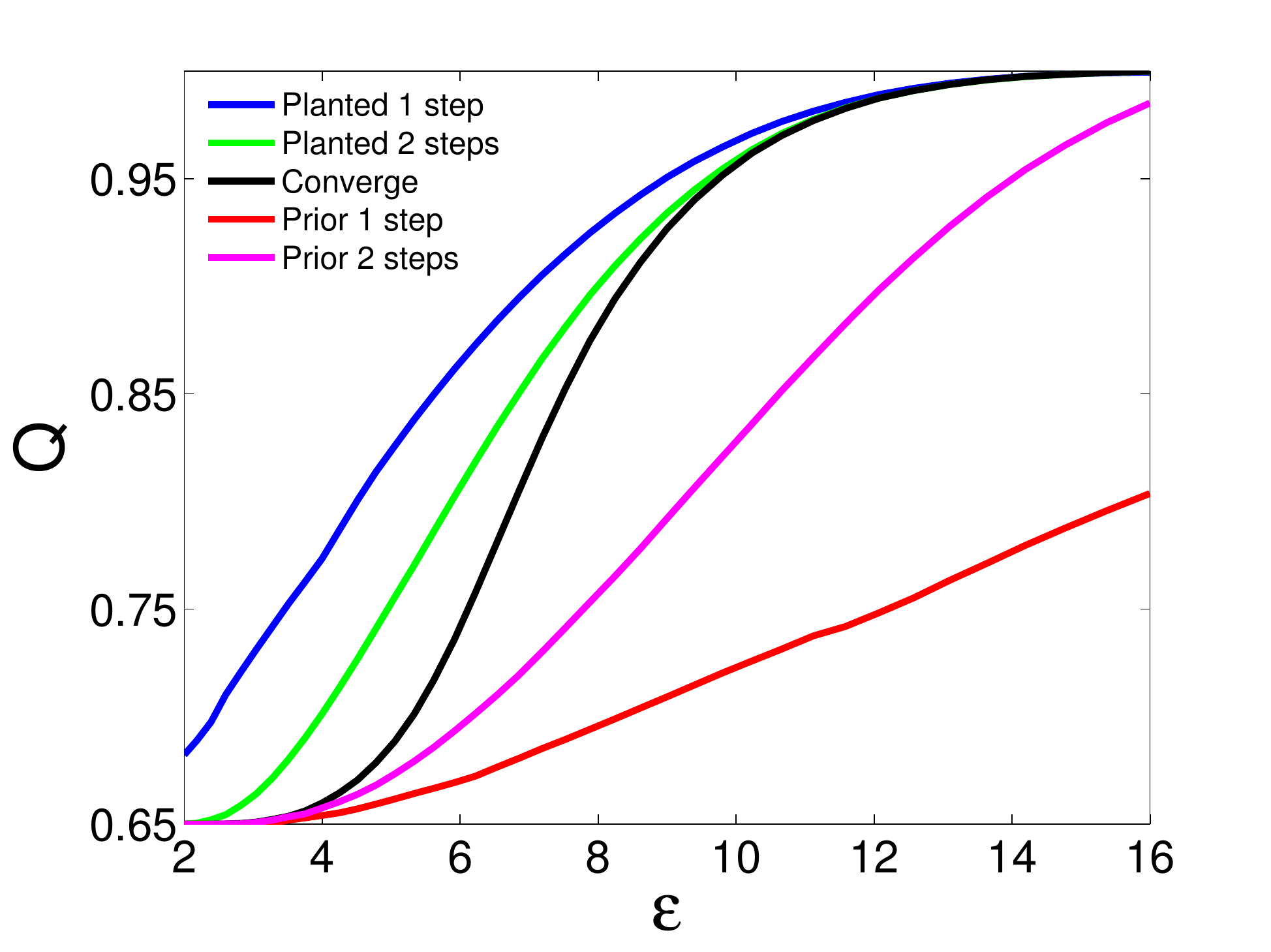}\qquad
\includegraphics[width=0.43\textwidth]{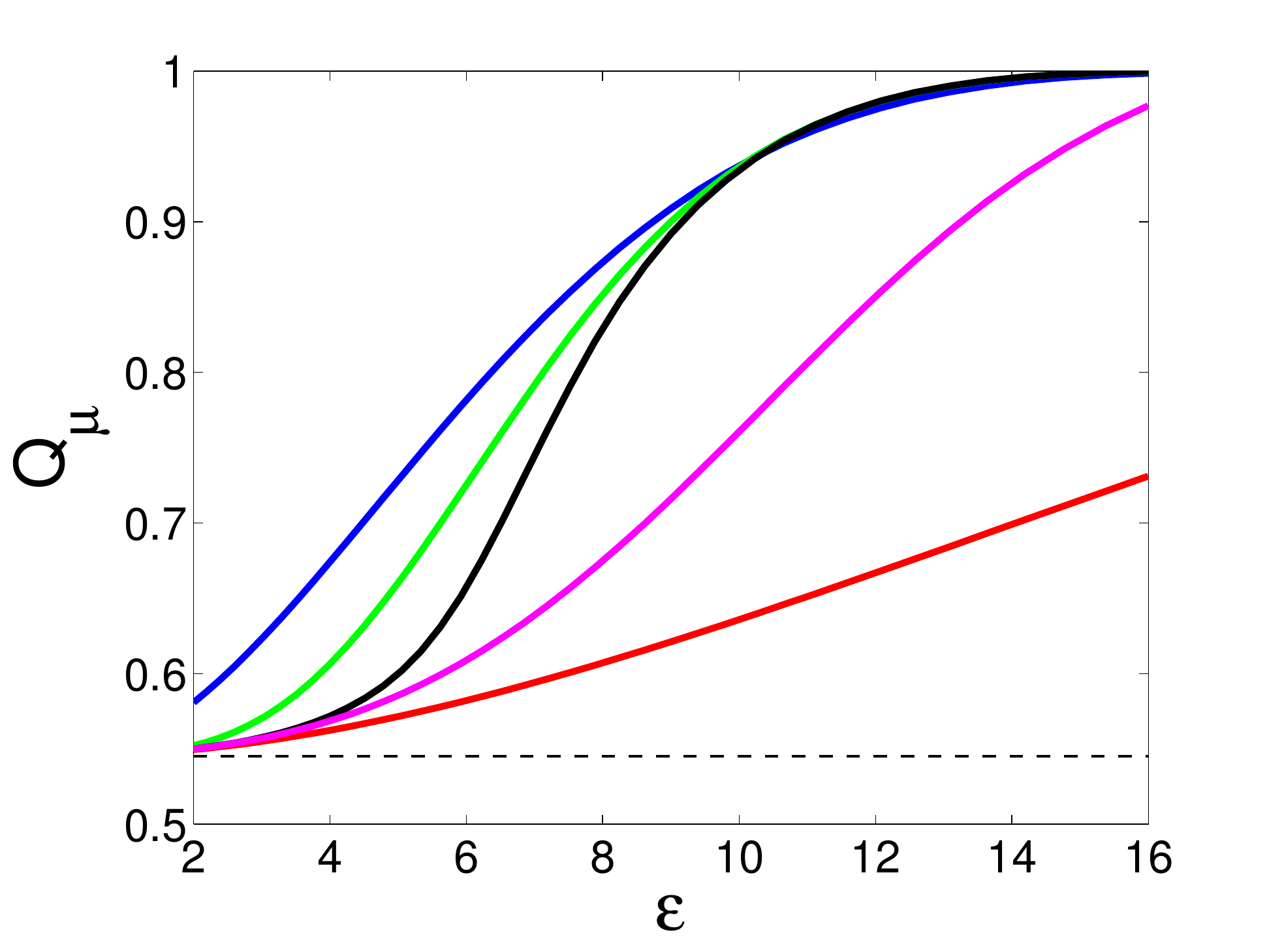}
\caption{The overlap $Q$ and marginal overlap $Q_\mu$ achieved by belief propagation on networks with $q=2$ groups, average degree $c=8$, and $\delta=0.6$, as a function of $\epsilon=\cin-\cout$.  The black curve in each panel shows the result of iterating belief propagation until it converges to a fixed point as in Fig.~\ref{fig:q2ovl}; the other curves show the overlap after $1$ and $2$ iterations.  The lower curves in each panel use initial messages $\mu_a^{i\to j} = \gamma_a$ equal to the prior probabilities of the groups as discussed in Section~\ref{sec:bpfinite}; the upper curves use initial messages $\mu_a^{i\to j} = \delta_{a,s_i}$ corresponding to the planted communities.  Each measurement is averaged over ten different networks of size $n=10^5$.}
\label{fig:q2steps}
\end{figure*}

In Fig.~\ref{fig:q2steps} we show the overlap~$Q$ (left) and marginal overlap~$Q_\mu$ (right) for $q=2$ groups.  In each panel, the black curve shows the overlap of the fixed point to which belief propagation converges if we continue iterating it.  Below that, we show two curves corresponding to iterating belief propagation for $t=1$ and $t=2$ steps where (as in Section~\ref{sec:bpfinite}) the messages are initially set equal to the prior probabilities $\mu_a^{i \to j} = \gamma_a$.  As we iterate, using information about the network from larger and larger neighborhoods, the accuracy of belief propagation improves and the curves approach the overlap for the fixed point from below.  We also show two further curves where the beliefs are initialized with the planted assignment $\mu_a^{i \to j} = \delta_{a,s_i}$ and these curves approach the overlap of the fixed point from above.  (The fixed point is the same for either initialization, since for $q=2$ and $\delta > 0$ there is no detectability transition.)

The curves with $t=2$, for either initialization, already give quite a good approximation to the final overlap when $\epsilon$ is either very low or very high.  Only in the intermediate region, close to the position of the detectability threshold for equal-sized groups (which in this case is at $\epsilon_c = 2\sqrt{8} \simeq 5.66$) is the approximation still poor after two iterations.  This agrees with previous observations that belief propagation converges quickly everywhere except in the vicinity of the transition~\cite{DKMZ11b}.  It also shows that, for $q=2$ groups, local information is enough to allow belief propagation to quickly approach optimal classification as the neighborhood radius increases.  (See~\cite{mossel-xu} for recent rigorous results on external fields or ``side information,'' showing that local algorithms also succeed in that setting.)

\begin{figure*}
\centering
\includegraphics[width=0.45\textwidth]{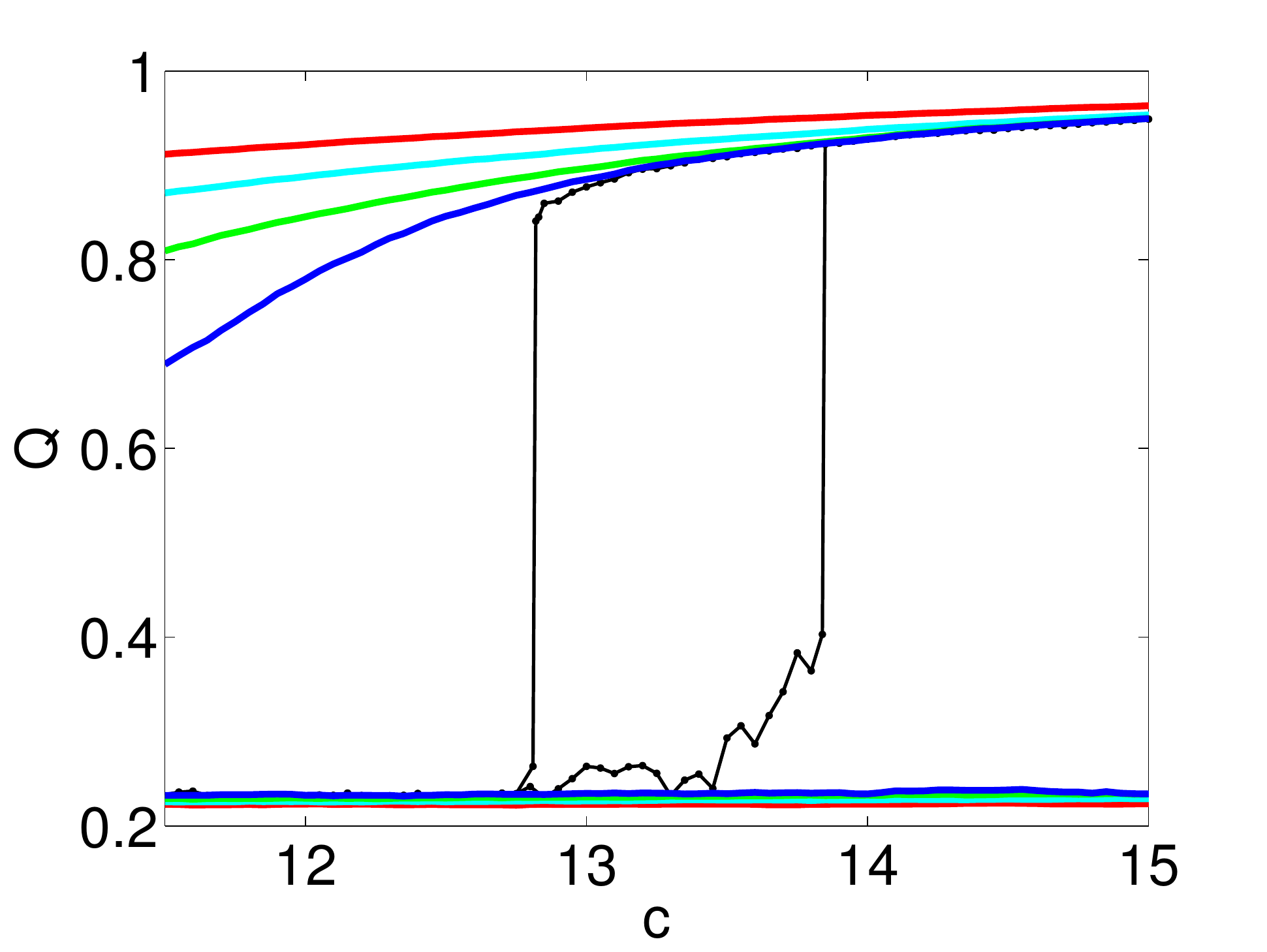}\qquad
\includegraphics[width=0.45\textwidth]{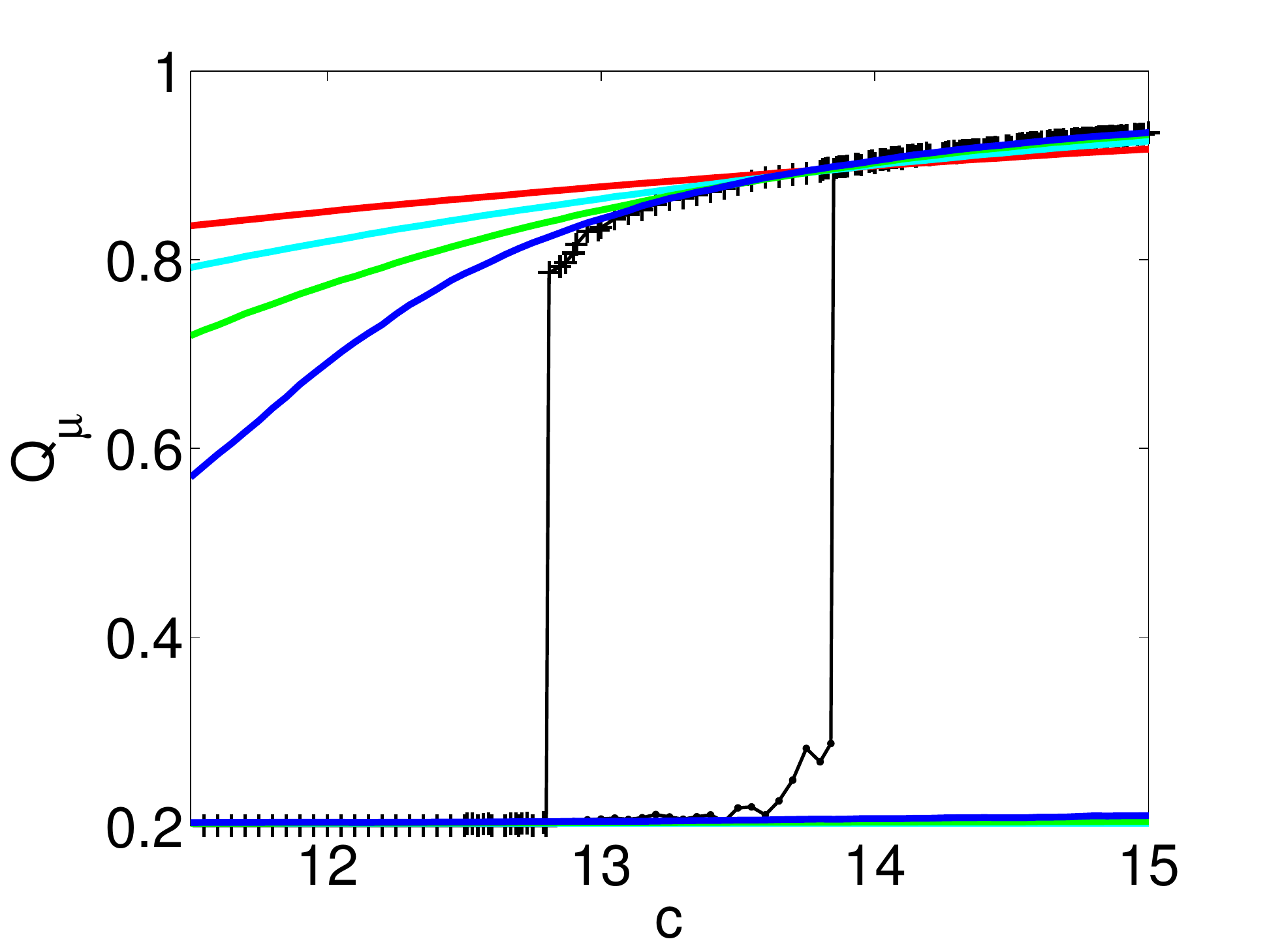}
\caption{The overlap $Q$ and marginal overlap $Q_\mu$ for belief propagation on fully disassortative networks generated by the stochastic block model with $q=5$ groups and $\delta=0.05$, as a function of the average degree~$c$.  The two black curves in each panel show the final converged result starting from the prior initialization and planted initialization respectively.  The other curves show the results of belief propagation after $t=1, 2, 4, 8$ iterations (red, cyan, green, blue curves respectively) starting from each initialization.  The converged result consists of one network at each point, and each measurement of the finite-step data is averaged over five different networks of size $n=10^5$.}
\label{fig:q5steps}
\end{figure*}

Figure~\ref{fig:q5steps} shows analogous results for $q=5$ groups in the fully disassortative case, for parameter values that encompass the coexistence region where both fixed points are stable.  In the coexistence region, initializing the beliefs with the prior probabilities---and thus using only local information---converges to the near-trivial fixed point, while initializing with the planted communities converges to the accurate fixed point.  This behavior is visible in the figure, with two lines in each panel (in black) showing the converged overlaps.  Outside the coexistence region these two lines agree but inside it they do not, showing that in this region there is a fundamental difference in the power of local vs.\ global information.

The remaining curves show the results for $t=1,2,4,8$ iterations.  With the prior initialization these results fail to register the first-order transition, instead following an analytic continuation of (an approximation to) the near-trivial fixed point.  Similarly, with the planted initialization we miss the spinodal transition and instead follow an approximate continuation of the accurate fixed point.  As a result, convergence from the ``wrong'' initialization to the final overlap is quite slow both above and below the coexistence region.

\section{Conclusions}
\label{sec:conclusions}

We have studied the detection of community structure in networks generated by the stochastic block model, a standard model of networks with well-defined clusters of nodes.  Previous studies have revealed the presence of a detectability transition in such networks, below which the communities are undetectable by any means.  In this paper we study the case where the symmetry between the groups is broken by having groups with unequal sizes or unequal average degrees.

We find that for the well-studied case of two groups, the detectability threshold disappears when the groups are unequal, making the accuracy a smooth function of the parameters of the model.  On the other hand, for $q > 4$ (or $q \ge 4$ in the disassortative case), where the detectability transition is first-order in the equal-groups case, the transition persists up to a certain amount of asymmetry.  Before this point is reached there is a coexistence between two competing solutions---one with low accuracy (but still better than chance) based on local information, and the other with higher accuracy based on global information.  As the amount of asymmetry increases, the coexistence region shrinks and finally disappears at a critical point, beyond which there is no sharp transition.  We conjecture that this local/global distinction may be a generic phenomenon in statistical inference whenever a symmetry is broken, both in networks and in other kinds of data.

\begin{acknowledgments}
  We are grateful to Florent Krzakala and Lenka Zdeborov\'a for helpful conversations.  This research was funded in part by the US National Science Foundation under grants DMS--1107796 and DMS--1407207 (MEJN) and by the John Templeton Foundation (PZ and CM).
\end{acknowledgments}

\bibliography{journals,zp,mark}

\end{document}